\documentclass[cernpreprint, atlasdraft=false, UKenglish, texmf, orcidlogo]{atlasdoc}
\usepackage[subfig=false, subfigure=true, subcaption=true]{atlaspackage}
\usepackage{atlasbiblatex}

\usepackage{atlasphysics}
\usepackage{letltxmacro}
\LetLtxMacro{\oldcite}{\cite}
\renewcommand{\cite}[1]{\mbox{\oldcite{#1}}}

\usepackage{multirow}
\usepackage{graphicx}

\graphicspath{{logos/}{figures/}}

\usepackage{ANA-STDM-2019-24-PAPER-defs}
\AtlasTitle{Measurement of the $W$-boson mass and width with the ATLAS detector using proton--proton collisions at $\sqrt{s}=7$ TeV}

\AtlasAbstract{%
Proton--proton collision data recorded by the ATLAS detector in 2011, at a centre-of-mass energy of 7~TeV, have been used for an improved determination of the $W$-boson mass and a first measurement of the $W$-boson width at the LHC. Recent fits to the proton parton distribution functions are incorporated in the measurement procedure and an improved statistical method is used to increase the measurement precision.
The measurement of the $W$-boson mass yields a value of
$m_W = 80366.5 \pm 9.8~(\text{stat.}) \pm 12.5~(\text{syst.})$~MeV~$= 80366.5 \pm 15.9$~MeV, and the width is measured as
$\Gamma_W = 2202 \pm 32~(\text{stat.}) \pm 34~(\text{syst.})$~MeV~$= 2202 \pm 47$~MeV. The first uncertainty components are statistical and the second correspond to the experimental and physics-modelling systematic uncertainties. Both results are consistent with the expectation from fits to electroweak precision data. The present measurement of $m_W$ is compatible with and supersedes the previous measurement performed using the same data.
}

\AtlasRefCode{STDM-2019-24}

\PreprintIdNumber{CERN-EP-2024-074}

\arXivId{2403.15085}

\AtlasJournalRef{Eur. Phys. J. C 84 (2024) 1309}
\AtlasDOI{10.1140/epjc/s10052-024-13190-x}

\hypersetup{pdftitle={ATLAS document},pdfauthor={The ATLAS Collaboration}}

\begin{document}

\maketitle

\tableofcontents

%
%

%
\section{Introduction}

At lowest order in the Standard Model~(SM) electroweak theory~\cite{GLASHOW1961579,SALAM1964168,PhysRevLett.19.1264} the $W$-boson mass, \mW, can be expressed solely as a function of the $Z$-boson mass, $m_Z$, the fine-structure constant, $\alpha$, and the Fermi constant, $G_\text{F}$. Higher-order corrections introduce an additional dependence of the $W$-boson mass on the gauge couplings and the masses of the heavy particles of the SM, such as the top-quark mass, $m_t$, and the Higgs boson mass, $m_H$~\cite{Sirlin:1980nh,Awramik:2003rn}. In extended theories, the loop corrections receive contributions from additional particles and interactions. The consistency of the SM and potential effects of new physics can therefore be probed by comparing the measured values of \mW with the results of global fits to the relevant physical parameters~\cite{deBlas:2021wap, Haller:2018nnx, Erler:1999ug}. The SM fit yields $\mW^\text{SM} = 80355 \pm 6$~\MeV~\cite{deBlas:2021wap, Haller:2018nnx}.
The present experimental situation is characterised by a significant tension between the precise measurement from the CDF Collaboration, $\mW = 80433.5 \pm 9.4$~\MeV~\cite{CDF:2022hxs}, and the average of the LEP~\cite{ALEPH:2013dgf}, D0~\cite{Abazov:2012bv}, ATLAS~\cite{STDM-2014-18} and LHCb~\cite{LHCb:2021bjt} measurements, $\mW = 80369.2 \pm 13.3$~\MeV~\cite{Amoroso:2023pey}.

The electroweak theory also predicts the total decay width of the $W$ boson, \GW. It is expected to be equal to the sum of the partial widths over three generations of lepton doublets and two generations of quark doublets, yielding an expected value of $\GW^\text{SM} = 2088\pm 1$~\MeV~\cite{deBlas:2021wap}. New particle candidates that couple to the $W$ boson and are lighter than \mW would open a new decay channel and alter \GW~\cite{Peskin:1991sw}. Examples are supersymmetric models in which the $W$ boson decays into the lightest super-partner of the charged gauge bosons and the lightest super-partner of the neutral gauge bosons~\cite{Heinemeyer:2013dia}. The current world average of $W$-boson width determinations yields a value of $\GW = 2085 \pm 42$~\MeV~\cite{ParticleDataGroup:2022pth}, and is based on measurements at LEP-2~\cite{ALEPH:2013dgf} and the Tevatron~\cite{Aaltonen:2012bp, D0:2013jba}. No measurement of \GW has been previously performed at the Large Hadron Collider~(LHC).

In this paper, an improved measurement of the $W$-boson mass as well as a first measurement of its width is presented, which is based on data from $\sqrt{s} = 7$~\TeV recorded by the ATLAS detector in 2011, i.e., the same data as was used for the first measurement of \mW at the LHC~\cite{STDM-2014-18}.
This was based on a $\chi^2$ considering statistical uncertainties only, where systematic uncertainties were included a posteriori through variations of the physics and calibration models within their uncertainties (the so-called `offset' method).

The present analysis uses an improved statistic based on the profile likelihood~(PLH)~\cite{Rolke:2004mj}. This technique performs a simultaneous determination of \mW together with a set of nuisance parameters describing the experimental and modelling uncertainties. The nuisance parameters are adjusted to optimally describe the data, yielding an overall improved model and some reduction in uncertainty compared with the fitting technique used previously.
With few sub-dominant exceptions, the sources of uncertainty considered in this measurement are either of experimental nature, or phenomenological, with model parameters derived from the data. A nuisance parameter representation is therefore adequate and the PLH technique can be applied.
The measurement of \GW relies on the same PLH statistic and on the same physics, detector and background model as those used for the determination of \mW.

Recent parton distribution functions~(PDFs) are studied within this work, and the dependence of the measurement results on the assumed PDF set is presented and discussed. The present analysis also aims at consolidating the earlier result of ATLAS, in the perspective of the latest measurement by CDF.


%
\section{ATLAS detector}
The ATLAS experiment~\cite{PERF-2007-01} is a multipurpose particle detector with a forward--backward symmetric
cylindrical geometry.
It consists of an inner tracking detector surrounded by a thin superconducting solenoid, electromagnetic and hadronic calorimeters,
and a muon spectrometer incorporating three large superconducting toroid magnets\footnote{ATLAS uses a right-handed coordinate system with its origin at the nominal interaction point~(IP)
in the centre of the detector and the \(z\)-axis along the beam pipe.
The \(x\)-axis points from the IP to the centre of the LHC ring,
and the \(y\)-axis points upwards.
Cylindrical coordinates \((r,\phi)\) are used in the transverse plane,
\(\phi\) being the azimuthal angle around the \(z\)-axis.
The pseudorapidity is defined in terms of the polar angle \(\theta\) as \(\eta = -\ln \tan(\theta/2)\).
Angular distance is measured in units of \(\Delta R \equiv \sqrt{(\Delta\eta)^{2} + (\Delta\phi)^{2}}\).}.

The inner-detector system~(ID) is immersed in a \SI{2}{\tesla} axial magnetic field
and provides charged-particle tracking in the range of $|\eta| < 2.5$. At
small radii, a high-granularity silicon pixel detector covers the
vertex region and typically provides three measurements per track. It is followed by the silicon
microstrip tracker, which usually provides
eight measurement points per track. These silicon
detectors are complemented by a gas-filled straw-tube transition
radiation tracker, which enables radially extended track
reconstruction within $|\eta| = 2.0$. The transition radiation tracker also provides electron
identification information based on the fraction of hits (typically 35
in total) above a higher energy-deposit threshold corresponding to
transition radiation.

The calorimeter system covers the pseudorapidity range $|\eta| < 4.9$.
Within the region $|\eta|< 3.2$, electromagnetic~(EM) calorimetry is
provided by high-granularity lead/liquid-argon~(LAr) calorimeters, with an additional thin LAr presampler covering $|\eta|<1.8$ to correct for upstream energy-loss fluctuations. The EM calorimeter is divided into a barrel section covering $|\eta|<1.475$ and two endcap sections covering $1.375<|\eta|<3.2$. For $|\eta|<2.5$, it is divided into three layers in depth, which are finely segmented in $\eta$ and $\phi$. Hadronic calorimetry is provided by a steel/scintillator-tile calorimeter, segmented into three barrel structures within $|\eta| = 1.7$ and two copper/LAr hadronic endcap calorimeters covering $1.5<|\eta|<3.2$. The solid-angle coverage is completed with forward copper/LAr and tungsten/LAr calorimeter modules in $3.1<|\eta|<4.9$, optimised for
electromagnetic and hadronic measurements, respectively.

The muon spectrometer~(MS) comprises separate trigger and
high-precision tracking chambers measuring the deflection of muons in a magnetic field generated by superconducting air-core toroids. The precision chamber system covers the region $|\eta| < 2.7$ with three layers of monitored drift tubes, complemented by cathode strip chambers in the forward region. The muon trigger system covers the range $|\eta| < 2.4$ with resistive plate chambers in the barrel, and thin gap chambers in the endcap regions.

A three-level trigger system is used to select events for offline analysis~\cite{PERF-2011-02}. The level-1 trigger is implemented in hardware and uses a subset of detector information to reduce the event rate to a design value of at most \SI{75}{\kHz}. This is followed by two software-based trigger levels that together reduce the event rate
to about \SI{300}{\Hz}.
A software suite~\cite{ATL-SOFT-PUB-2021-001} is used in data simulation, in the reconstruction
and analysis of real and simulated data, in detector operations, and in the trigger and data acquisition
systems of the experiment.


%
\section{Measurement overview and analysis strategy} %

\subsection{Data samples and event simulation\label{sec:data}}

The data sample consists of $W\to e\nu$ and $W\to \mu\nu$ candidate events, collected in 2011 with the ATLAS detector in proton--proton collisions at the LHC, at a centre-of-mass energy of $\sqrt{s}=7$~\TeV. The data collected with all relevant detector systems operational correspond to approximately $4.6$~fb$^{-1}$ and $4.1$~fb$^{-1}$ of integrated luminosity in the electron and muon channels, respectively.

The \POWHEG Monte Carlo~(MC) generator~(v1/r1556)~\cite{Nason:2004rx,Frixione:2007vw,Alioli:2010xd}
is used for the simulation of $W$- and $Z$-boson production and decay in the electron, muon, and $\tau$-lepton channels, and is interfaced to \PYTHIA[8]~(v8.170) for the modelling of the parton shower, hadronisation, and underlying event~\cite{Sjostrand:2006za-corr,Sjostrand:2007gs}. Parton shower and underlying event parameters are set according to the AZNLO tune~\cite{STDM-2012-23}. The CT10 PDF set~\cite{Lai:2010vv} is used for the hard process, and the CTEQ6L1 PDF set~\cite{Pumplin:2002vw} is used in the parton shower. The $Z$-boson simulation includes the effect of virtual photon exchange. The $W$- and $Z$-boson rapidity and \pt distributions are reweighted to optimise the description of the data, as described in Section~\ref{sec:modunc}. The change in the final-state distributions from updating the distributions to more recent PDFs is evaluated using \POWHEG.

QED final-state radiation~(FSR) is simulated using \PHOTOS~(v2.154)~\cite{Golonka:2005pn}. Decays of $\tau$-leptons are handled by \PYTHIA[8], taking into account polarisation effects. The $W$- and $Z$-boson event yields are normalised according to their measured cross sections, and the experimental uncertainties of 1.8\% and 2.3\% are assigned to the $W^{+}/Z$ and $W^{-}/Z$ production cross-section ratios, respectively~\cite{STDM-2012-20}. The $W$-boson production samples assume $\mW = 80399$~\MeV and $\GW = 2085$~\MeV.

Background processes such as top-quark pair and single-top-quark production are modelled using the \MCatNLO MC generator~(v4.01)~\cite{Frixione:2002ik-corr,Frixione:2003ei-corr,Frixione:2005vw}, interfaced to \HERWIG and \JIMMY for the parton shower. Gauge-boson pair production ($WW$, $WZ$, $ZZ$) is simulated with \HERWIG~(v6.520). The CT10 PDF set is used in all these samples.

The response of the ATLAS detector is simulated using a software suite~\cite{SOFT-2010-01} based on \GEANT~\cite{Agostinelli:2002hh-corr}. The hard-scattering process is overlaid with additional proton--proton interactions, simulated with \PYTHIA[8]~(v8.165) using the A2 tune~\cite{ATL-PHYS-PUB-2012-003}. The distribution of the average number of interactions per bunch crossing $\avg{\mu}$ spans the range $2.5$--$16.0$, with a mean value of approximately $9.0$.

\subsection{Selection of electrons and muons and reconstruction of the recoil}

Object definitions are unchanged compared to Ref.~\cite{STDM-2014-18}. Electron candidates are reconstructed from clusters of energy deposited in the electromagnetic calorimeter and associated with at least one track in the ID~\cite{PERF-2013-03,PERF-2013-05}. Quality requirements are applied to the associated tracks in order to reject poorly reconstructed charged-particle trajectories. The energy of the electron is reconstructed from the energy collected in calorimeter cells within an area of size $\Delta \eta \times \Delta \phi = 0.075 \times 0.175$ in the barrel, and $0.125 \times 0.125$ in the endcaps. The energy measurement relies on a multivariate regression algorithm developed and optimised on simulated events. The kinematic properties of the reconstructed electron are inferred from the energy measured in the EM calorimeter and from the pseudorapidity and azimuth of the associated track. Electron candidates are required to fulfil tight identification requirements~\cite{PERF-2013-03}, and their transverse momentum, \ptl, and pseudorapidity should satisfy $\ptl > 15~\GeV$ and $|\eta|<2.4$. As in the previous result, the pseudorapidity range $1.2<|\eta|<1.8$ is excluded from the measurement. Background from jets misidentified as electrons is reduced using additional isolation requirements using the activity in the ID and calorimeter nearby the electron candidates passing the kinematic and identification selections~\cite{STDM-2014-18}.

Muon candidates are reconstructed independently in the ID and in the MS, and a combined muon candidate is formed from the statistical combination of the ID and MS track parameters~\cite{PERF-2014-05}. The kinematic properties of the reconstructed muon are defined using the ID track parameters alone, which allows a simpler calibration procedure. Muon candidates are required to have $\ptl>20~\GeV$ and $|\eta|<2.4$~\cite{STDM-2014-18}. Similarly to the electrons, the multijet background is reduced by applying an isolation requirement~\cite{STDM-2014-18}.

The recoil, $\utvec$, is an estimator of the $W$- or $Z$-boson transverse momentum. It is reconstructed from the vector sum of the transverse energy of all clusters measured in the calorimeters, excluding clusters located at a distance $\Delta R < 0.2$ from electron or muon candidates. The definition of $\utvec$ does not involve the explicit reconstruction of jets to avoid possible $\pt$ threshold effects.

\subsection{$W$-boson kinematics and event selection}

The transverse momentum vector of charged leptons from the $W$-boson decay, $\vec{p}_{\text{T}}^{\,\ell}$, is measured as summarised in the previous section. The transverse momentum of the decay neutrino is inferred from the missing transverse momentum vector, $\mptvec$, defined as $ \mptvec = -\left(\vec{p}_{\text{T}}^{\,\ell} + \utvec\right)$. The $W$-boson transverse mass, \mt, is derived from $\mpt$ and from the transverse momentum of the charged lepton as $ \mt = \sqrt{2 \pt^\ell \mpt (1-\cos{\Delta\phi})}$, where $\Delta\phi$ is the azimuthal opening angle between the charged lepton and the missing transverse momentum.

The $W$-boson sample is collected using triggers requiring at least one muon candidate with transverse momentum larger than $18~\GeV$ or at least one electron candidate with transverse momentum larger than $20~\GeV$. The transverse-momentum requirement for the electron candidate was raised to $22~\GeV$ in later data-taking periods to cope with the increased instantaneous luminosity delivered by the LHC\@. Selected events are required to have a reconstructed primary vertex with at least three associated tracks.

The sample of $W$-boson candidate events is selected by requiring exactly one reconstructed electron or muon candidate with $\ptl > 30~\GeV$. The leptons are required to match the corresponding trigger signal. The magnitude of the recoil is required to satisfy $\ut < 30~\GeV$, the missing transverse momentum $\mpt > 30~\GeV$ and the transverse mass $\mt > 60~\GeV$. Approximately $5.89 \times 10^6$ candidate events are selected in the $W\to e\nu$ channel and $7.84 \times 10^6$ events in the $W\to\mu\nu$ channel.

\subsection{$W$-boson mass analysis updates\label{sec:updates}}

The selected $W$-boson event sample includes events from various background processes. Background contributions from $Z$-boson, $W\rightarrow\tau\nu$, boson pair, and top-quark production are estimated using simulation, and represent about 6.4\% of the total sample in the muon channels, and about 3.1\% in the electron channels. Contributions from multijet production are estimated with data-driven techniques, and are detailed in Ref.~\cite{STDM-2014-18}. Compared with that reference, the multijet background yield was re-evaluated using the final luminosity calibration for Run 1~\cite{DAPR-2011-01}, resulting in a 1--2\% decrease of the contamination of multijet background control regions by electroweak processes, and a corresponding 20\% increase in the estimated multijet background yield in the electron channel. This background now represents 1.2\% of the total sample, and agrees with the previous measurement within uncertainties. The multĳet background in the muon channel is unaffected, due to the smaller contamination in this case. In addition, uncertainties in the multijet distributions were previously propagated to the \mW measurement through fluctuations of the extrapolation parameters; an eigenvector decomposition is used in the present analysis.

The results of Ref.~\cite{STDM-2014-18} were obtained using the CT10nnlo PDF set and compared with results using the CT14~\cite{Dulat:2015mca} and MMHT2014 PDF sets~\cite{Harland-Lang:2014zoa}. The present analysis extends the study of the PDF dependence of the fit results to the ATLASpdf21~\cite{STDM-2020-32}, CT18, CT18A~\cite{Hou:2019efy}, MSHT20~\cite{Bailey:2020ooq}, NNPDF3.1~\cite{NNPDF:2017mvq} and NNPDF4.0~\cite{NNPDF:2021njg} sets.

In the previous measurement, \mW was determined with the $W$-boson width fixed to the SM prediction. In the present analysis, this assumption is relaxed by treating \GW as a source of systematic uncertainty, considering the SM value and uncertainty of $\GW^\text{SM} = 2088 \pm 1$~\MeV. The $W$-boson width is also extracted assuming the SM prediction and uncertainty of the $W$-boson mass, $\mW^\text{SM}= 80355 \pm 6$~\MeV.

Finally, an improved statistic is used for the fit as described in the following section.

\subsection{Statistical analysis\label{sec:fitting}}

The previous measurement used separate template fits to the \ptl and \mt distributions observed in different event categories. The $W$-boson candidate events were classified according to the charge, flavour and pseudorapidity of the final state lepton, as summarised in Table~\ref{tab:Categories}. In the fit, the $\chi^2$ of the comparison between data and simulation was minimised considering statistical uncertainties only; systematic uncertainties were included by varying the parameters determining the templates within their uncertainties, and repeating the fits.

\begin{table}[tbp]
\centering\caption{\label{tab:Categories} Summary of the 28 categories and kinematic distributions used in the \mW measurement for the electron and muon decay channels.}\begin{tabular}{lcc}
\toprule
Decay channel		&	$W\rightarrow e\nu$				&	$W\rightarrow \mu\nu$ \\
\midrule
Kinematic distributions &	$\pt^\ell$, \mt					&	$\pt^\ell$, \mt \\
Charge categories	&	$W^+$, $W^-$					&	$W^+$, $W^-$ \\
$|\eta_\ell|$ categories	&	$[0,0.6]$, $[0.6,1.2]$, $[1.8,2.4]$	& $[0,0.8]$, $[0.8,1.4]$, $[1.4,2.0]$, $[2.0,2.4]$ \\
\bottomrule
\end{tabular}
\end{table}

The present analysis performs a simultaneous optimisation of \mW or \GW, and of nuisance parameters describing systematic uncertainties, through a global profile likelihood fit in all event categories for a given kinematic distribution. The likelihood function, which describes how compatible with each other the data and MC distributions are, is given by

\begin{equation}\label{eq:plh}
{\cal L} \left( \vec n \right | \mu,\vec\theta) =
\prod\limits_{j} \prod\limits_{i} \text{Poisson} \left(n_{ji} | \nu_{ji}(\mu,\vec\theta) \right)\cdot \text{Gauss}\left(\vec\theta\right),
\end{equation}

where $\vec n$ represents the observed distributions in data, and $n_{ji}$ is the number of events observed in data in bin $i$ of the distribution in a given category $j$. It is the input to the Poisson distribution with expectation $\nu_{ji}(\mu,\vec\theta) = S_{ji}(\mu,\vec\theta)+B_{ji}(\vec\theta)$, of $S_{ji}$ events from signal and $B_{ji}$ events from background contributions. The parameter of interest, $\mu$, represents variations in \mW or \GW with respect to a conventional reference, $\mu_\text{ref}$. Uncertainties of the signal and background distributions are encapsulated as nuisance parameters~(NPs), denoted by $\vec\theta$ in Eq.~\eqref{eq:plh}, for which a normal probability distribution is assumed. The expected number of events \(v_{ji}\) is parameterised as
\begin{equation}
\begin{aligned}
v_{ji}(\mu,\vec\theta)&=\Phi\times\left[S_{ji}^{\text{nom}}+\mu\times\left(S_{ji}^\mu-S_{ji}^{\text{nom}}\right)\right]+\sum_{s}\theta_{s}\times\left(S_{ji}^{s}-S_{ji}^{\text{nom}}\right) \\%
&+B_{ji}^{\text{nom}}+\sum_{b}\theta_{b}\times\left(B_{ji}^{b}-B_{ji}^{\text{nom}}\right),        \end{aligned}
\label{eq:sigplh}
\end{equation}

where \(\Phi\) is an overall, unconstrained normalisation factor ensuring that the total $W^\pm$ signal rate always adjusts to the number of events in data, \(S_{ji}^{\text{nom}}\) and \(B_{ji}^{\text{nom}}\) are the nominal distributions of signal and background, respectively, while $s$ and $b$ represent nuisance parameters acting on signal and background contributions.

\begin{figure}[tbp]
\centering
\subfigure[]{\includegraphics[width=0.49\textwidth]{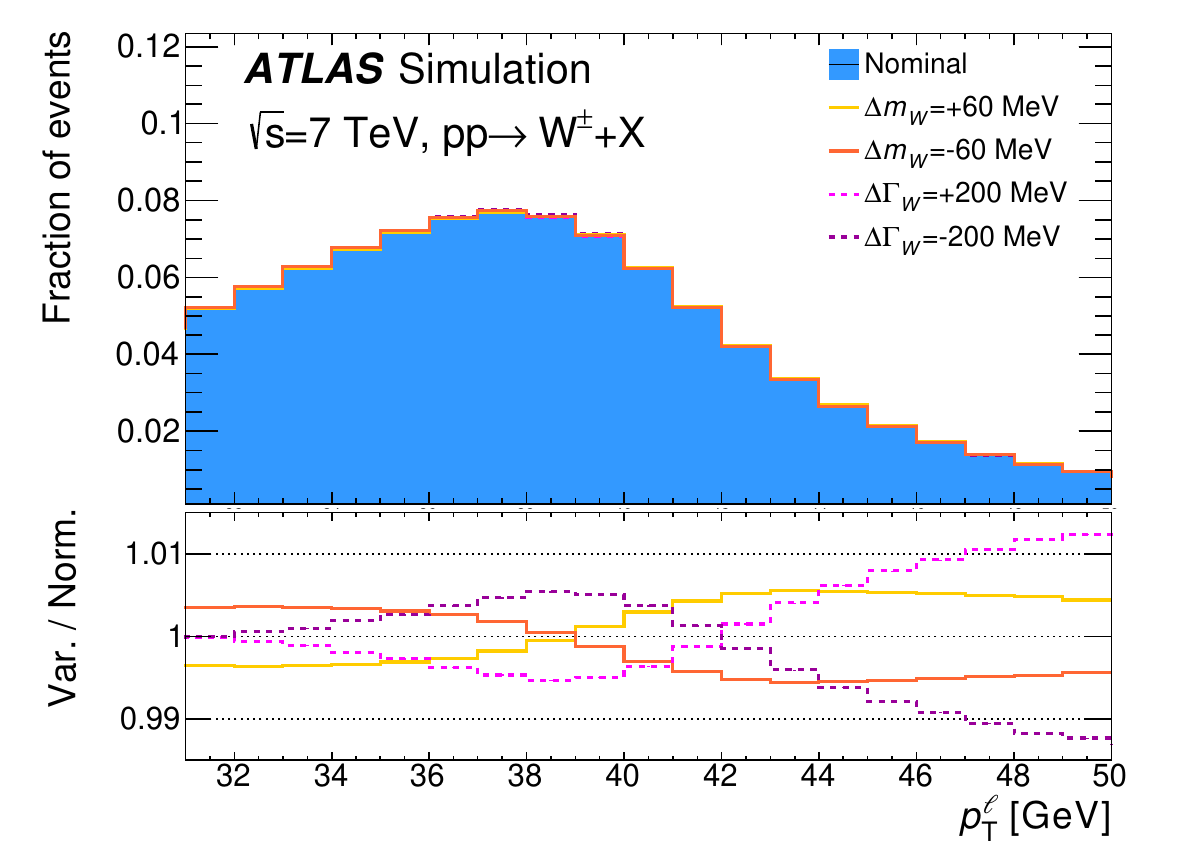}}
\subfigure[]{\includegraphics[width=0.49\textwidth]{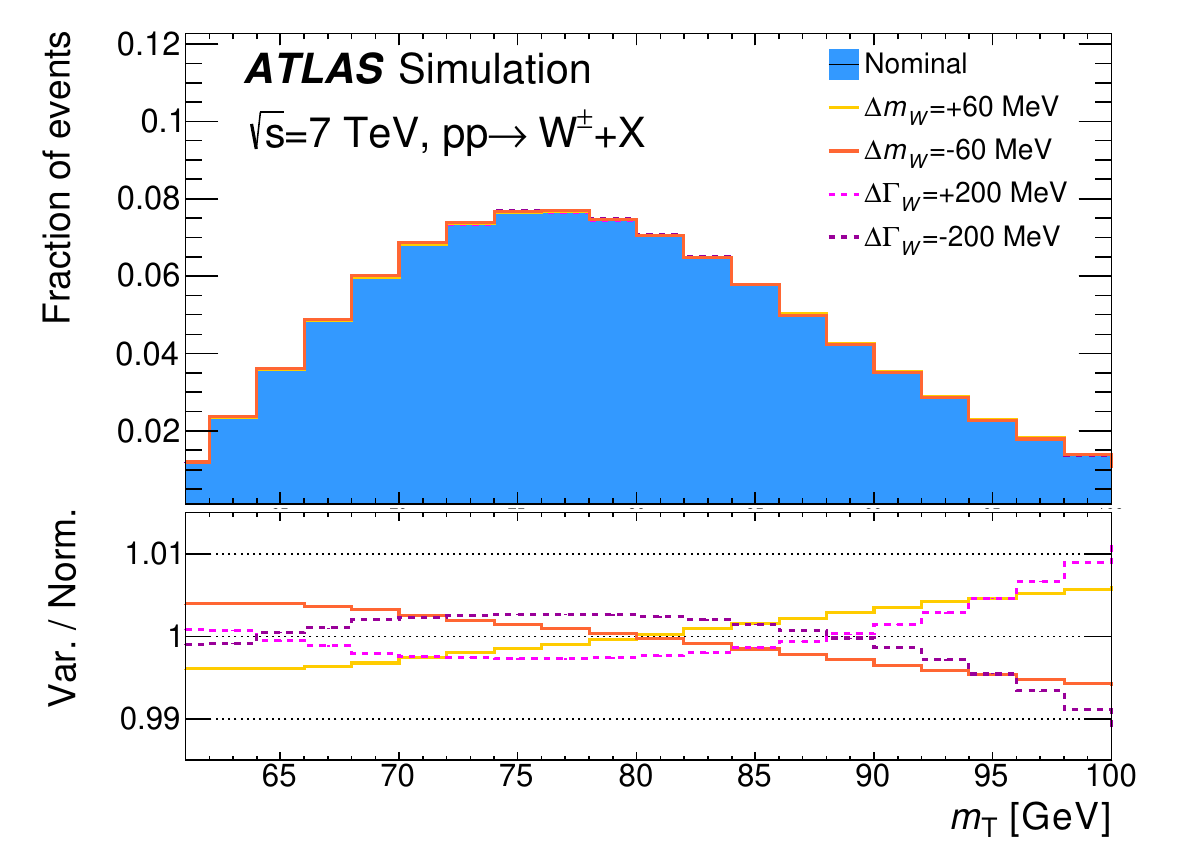}}
\caption{Simulated kinematic distributions of (a) \ptl and (b) \mt in $W^\pm \to \mu^\pm\nu$ events, for $W$-boson mass and width values of $\mW = 80399$~\MeV and $\GW = 2085$~\MeV. The ratio panels represent the relative effect of varying these parameters by $\pm 60$~\MeV and $\pm 200$~\MeV, respectively.}
\label{fig:mwgwtemplates}
\end{figure}

Changes in $\mu$ and $\vec\theta$ lead to changes in the expected signal and background distributions, which are interpolated using a polynomial morphing procedure. Signal templates for arbitrary values of \mW or \GW are obtained from the same simulation sample through a reweighting of the $W$-boson Breit--Wigner distribution. Templates representing systematic variations are determined from two-sided one- and two-sigma variations of the corresponding sources of uncertainty. The effect of varying \mW or \GW on the \ptl and \mt distributions is illustrated in Figure~\ref{fig:mwgwtemplates}. The procedure to interpolate between these points during the PLH fit was extensively tested, and excellent closure was observed.

As the new fitting method allows to better optimise the total uncertainty of the measurement due to the inclusion of NPs in the likelihood, the nominal fit ranges for the \mW measurement are re-evaluated. The updated optimal fit ranges are $30 < \ptl < 50$~\GeV and $60 < \mt < 100$~\GeV, in contrast with $32 < \ptl < 45$~\GeV and $66 < \mt < 99$~\GeV used in the previous measurement. For the determination of \GW, the same ranges are used as in the \mW measurement.

The baseline results rely on a numerical minimisation of the likelihood from Eq.~\eqref{eq:sigplh}. For ancillary studies, such as the decomposition of uncertainties, fit range variations, and to estimate the correlation between the \mt and \ptl fits, the following assumptions are made: in the limit where all uncertainties are Gaussian and the dependence of $\nu_{ji}(\mu,\vec{\theta})$ on $\mu$ and $\vec{\theta}$ is linear, the likelihood can be written as
\begin{equation}
\begin{aligned}
-2\ln{\cal L} \left(  \vec{n}\right | \mu,\vec{\theta} ) =
\sum_{j} \sum_{i} \left( \frac{ n_{ji} - \nu_{ji}(\mu_\text{ref},\vec{0}) - \frac{\partial\nu_{ij}}{\partial\mu}(\mu-\mu_\text{ref}) - \sum_{t} \frac{\partial\nu_{ij}}{\partial\theta_t}\theta_t}{\sigma_{ji}} \right)^2
+ \sum_{t} \theta_t^{2},
\end{aligned}
\end{equation}
and the minimisation and uncertainty estimation can be performed analytically~\cite{Pinto:2023yob}. This approach gives results within 2~\MeV from the nominal fits and is much faster.

The decomposition of the post-fit uncertainties is performed according to the methods of Ref.~\cite{Pinto:2023yob}. The uncertainty components are defined to represent the contribution of the pre-fit uncertainty in the corresponding sources to the total uncertainty of the measurement, consistently with standard error propagation.


\section{Experimental corrections and uncertainties} %

The \ptl and \mt distributions are affected by the lepton energy calibration and by the calibration of the recoil. Lepton momentum corrections are derived exploiting $Z\rightarrow\ell\ell$ event samples and the precisely measured value of $m_Z$~\cite{ALEPH:2005ab}, and the recoil response is calibrated using the expected momentum balance between \ut and $p_\text{T}^{\ell\ell}$~\cite{STDM-2014-18}. Lepton identification and reconstruction efficiency corrections are determined from $W$- and $Z$-boson events using the tag-and-probe method~\cite{PERF-2013-03,PERF-2014-05}.

A precision on the energy and momentum scale for electrons and muons of $O(10^{-4})$ is achieved, with somewhat larger uncertainty for the muons in the high-$\eta$ region. The response and resolution of $\ut=|\vec \ut|$ is determined with a precision of a few percent. The experimental precision is limited by the finite size of the $Z$-boson sample, and by systematic uncertainties in the modelling of the distributions used in the calibration procedures.

\subsection{Uncertainty propagation}

Systematic uncertainties in the determination of \mW and \GW are evaluated by varying the calibration model parameters within their uncertainty. For two-sided systematic uncertainties, separate templates are produced for 68\% confidence level~(CL) upwards and downwards variations. Systematic uncertainties that are estimated independently in many kinematic bins are propagated through simultaneous random variations of the corresponding parameters within their uncertainty and generating templates for each variation. A principal component analysis~(PCA)~\cite{Pearson, Hotelling} is used to transform these variations into a set of uncorrelated two-sided uncertainties, preserving the total uncertainty. This approach is used for the statistical uncertainties of the electron and muon efficiencies, as well as for the recoil calibration, and allows a faithful representation of these uncertainties using a reduced set of nuisance parameters.

Because of the finite size of the MC samples, some systematic variations contribute significant statistical fluctuations in the final-state distributions. A smoothing procedure is applied to remove such fluctuations, preserving the normalisation of each variation. For calibration systematic uncertainties, the effect of the upwards and downwards variations are symmetrised. The impact of the smoothing and symmetrisation procedures on the best-fit values and uncertainties are below 1 and 0.1~\MeV, respectively.

The effect of each systematic variation is decomposed into corresponding uncertainties in the normalisation and in the shape of the final state distributions. Systematic uncertainties that yield differences smaller than 0.01\% in the normalised \ptl distribution and smaller than 0.02\% in the normalised \mt distribution are removed, reducing the number of shape systematic variations by a factor of two. The change in the total measurement uncertainty is less than 1\% of itself, while central values change by less than 0.1~\MeV. This pruning procedure simplifies the likelihood, stabilising and accelerating the fit procedure.

\subsection{Sources of uncertainty}

The electron calibration and selection efficiencies account for 75 sources of uncertainty in the \ptl distribution and 58 in the \mt distribution, including the energy scale and resolution as well as the electron identification, isolation, and trigger efficiencies. Of these uncertainties, 23 originate from the energy calibration and are treated as two-sided systematic uncertainties, while 52~(\ptl) and 35~(\mt) systematic variations come from the trigger, reconstruction, identification and isolation efficiencies. PCA is utilised to handle those systematic variations. Similarly, the muon response and efficiencies contribute 83~(\ptl) and 76~(\mt) sources of uncertainty, of which 6 are treated as two-sided uncertainties.

The calibration of the hadronic recoil yields 36 sources of systematic uncertainties for the \mt distributions, but only 7 sources for the \ptl distributions since the impact on the latter is only due to the hadronic recoil requirement in the signal selection. Of these uncertainties, 3 are two-sided uncertainties, while 4~(\ptl) and 33~(\mt) PCA variations are taken into account.


\section{Physics corrections and uncertainties} %

\subsection{Electroweak uncertainties\label{sec:ewuncert}}

The dominant source of electroweak corrections to $W$-boson production originates from QED final-state radiation, and is simulated with \PHOTOS. The effect of QED initial-state radiation~(ISR) is also included through the \PYTHIA[8] parton shower~(PS). Other sources of electroweak corrections are not included in the simulated event samples, and their full effects are considered as systematic uncertainties. Systematic uncertainties from missing higher-order electroweak corrections are estimated considering the same sources of uncertainty as in Ref.~\cite{STDM-2014-18}. An improvement of the present analysis is that the corresponding uncertainties are evaluated at detector level instead of generator level, which was a simplification used in the previous analysis as these are not leading uncertainties. The detector-level systematic variations are obtained by applying detector response and efficiency migration matrices derived from samples of simulated signal events described in Section~\ref{sec:data}. Their impact on \mW is larger than for generator-level variations by typically 20\%.

\subsection{QCD model and uncertainties \label{sec:modunc}}

The rapidity, transverse momentum and decay distributions of the simulated $W$- and $Z$-boson samples are reweighted to include the effects of higher-order QCD corrections, which improves the agreement between the data and simulation. The differential cross section as a function of the boson rapidity, $\text{d}\sigma(y)/\text{d}y$, and the coefficients describing angular distributions of decay leptons, $A_i$~\cite{STDM-2014-10}, are calculated at $O(\alpha_\text{s}^2)$ in fixed-order QCD\@. The transverse-momentum spectrum at a given rapidity, $\text{d}\sigma(\pt,y)/(\text{d}\pt\,\text{d}y) \cdot (\text{d}\sigma(y)/\text{d}y)^{-1}$, is modelled using the \PYTHIA[8] MC generator, with parameters adjusted to reproduce the measured $Z$-boson \pt distribution at $\sqrt{s}=7$~\TeV~\cite{STDM-2012-23}. The resulting tune, called AZ in the following, predicts $W$-boson \pT distributions that agree with measurements at $\sqrt{s}=5.02$~and 13~\TeV~\cite{ATLASpTW}.

PDF uncertainties are calculated for the CT10, CT14, CT18, CT18A, MMHT2014, MSHT20, NNPDF3.1, NNPDF4.0 and ATLASpdf21 sets using the Hessian method~\cite{Pumplin:2001ct}, where each eigenvector of the PDF fit covariance matrix defines a pair of PDF uncertainty variations and a corresponding nuisance parameter in the PLH fit. The CT10, CT14, CT18 and CT18A variations correspond to 90\%~CL, and are rescaled to match the 68\%~CL\@. PDF uncertainty variations are constrained to leave the predicted \ptz distribution unchanged, propagating only the part of the PDF uncertainty in the \ptw distribution that is uncorrelated to \ptz~\cite{STDM-2014-18}. To achieve this, the impact of each PDF eigenvector on the \ptw and \ptz distributions is calculated, and the corresponding uncertainty in the \ptw distribution is defined from the ratio of the varied \ptw and \ptz distributions. This procedure is equivalent to performing an explicit parton shower tune for each PDF variation but simpler in practice. The uncertainties in the AZ tune parameters are propagated separately as described below.

The \PYTHIA[8] parton shower model contributes additional sources of uncertainty in the \ptw distribution. The AZ tune parameters are assumed universal between $Z$- and $W$-boson production, and their uncertainties are propagated to the $W$-boson final-state distributions. The initial-state charm and bottom quark masses affect the \pt spectrum, and the corresponding uncertainties are estimated by varying their respective masses by $\pm 0.5\,\GeV$ and $\pm 0.8\,\GeV$, respectively. Uncertainties in the shower evolution are parameterised through variations of the factorisation scale, $\mu_\text{F}$, by factors of $0.5$ and $2.0$ with respect to the central choice $\mu_\text{F}^2 = p_{\text{T},0}^2 + \pt^2$, where $p_{\text{T},0}$ is an infrared cut-off, and \pt{} is the evolution variable of the parton shower~\cite{Sjostrand:2004ef}. The variations are applied independently to the light-quark, charm-quark and bottom-quark-induced processes, and are propagated considering only the relative impact on the $\pt^W$ and $\pt^Z$ distributions, as done for the PDF uncertainties. Differences between the \PYTHIA[8] and \HERWIG[7] predictions for this ratio were found to be negligible.

The accuracy of the next-to-next-to-leading-order~(NNLO) predictions for the angular coefficients $A_0$--$A_7$ is validated by comparing to the corresponding measured values in $Z$-boson production~\cite{STDM-2014-10}. The $Z$-boson data uncertainties are propagated to the $W$-boson predictions, which assumes that NNLO predictions have similar accuracy for the $W$- and $Z$-boson processes, and are validated within the experimental precision of the $Z$-boson data. The observed disagreement between data and prediction for the $A_2$ coefficient is taken as additional uncertainty. Similarly to some experimental uncertainties, random angular coefficient variations are treated with a PCA to produce uncorrelated two-sided uncertainties.

The effect of missing higher-order corrections on the NNLO predictions of the normalised rapidity distributions and the effect of the LHC beam-energy uncertainty of 0.65\% were both found to be negligible.


%
\section{Improved measurement of the $W$-boson mass} %

The improvements to the previous \mW measurement described in Sections~\ref{sec:updates} and~\ref{sec:fitting} are implemented in several steps.
The impact of the analysis updates is evaluated using the same statistical method as in the previous measurement, and yields a change of the measured $W$-boson mass from $80369.5 \pm 18.5$~\MeV to $80371.9 \pm 18.7$~\MeV, corresponding to a shift of 2.4~\MeV and a minimal increase of the total uncertainty. The analysis is then repeated using the CT10nnlo PDF set and unchanged systematic uncertainties but implementing the PLH approach. This provides a test of the stability of the measurement under the change of statistic used in the fit. The final step consists of updating the measurement to more recent PDF sets, one of which is used to define a new baseline.

\subsection{Results with CT10nnlo and consistency tests}

\begin{figure}[tbp]
\centering
\subfigure[]{\includegraphics[width=0.49\textwidth]{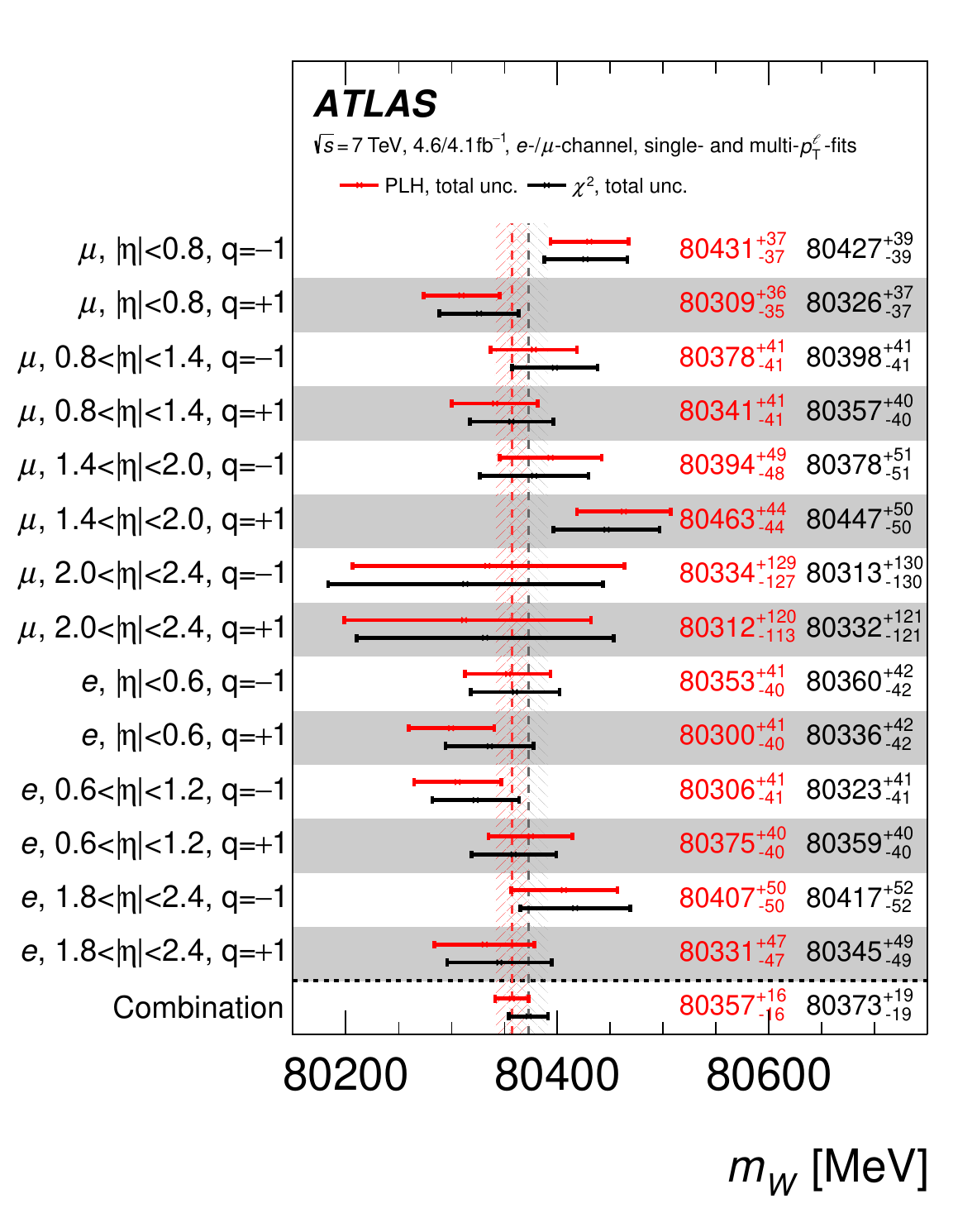}}
\subfigure[]{\includegraphics[width=0.49\textwidth]{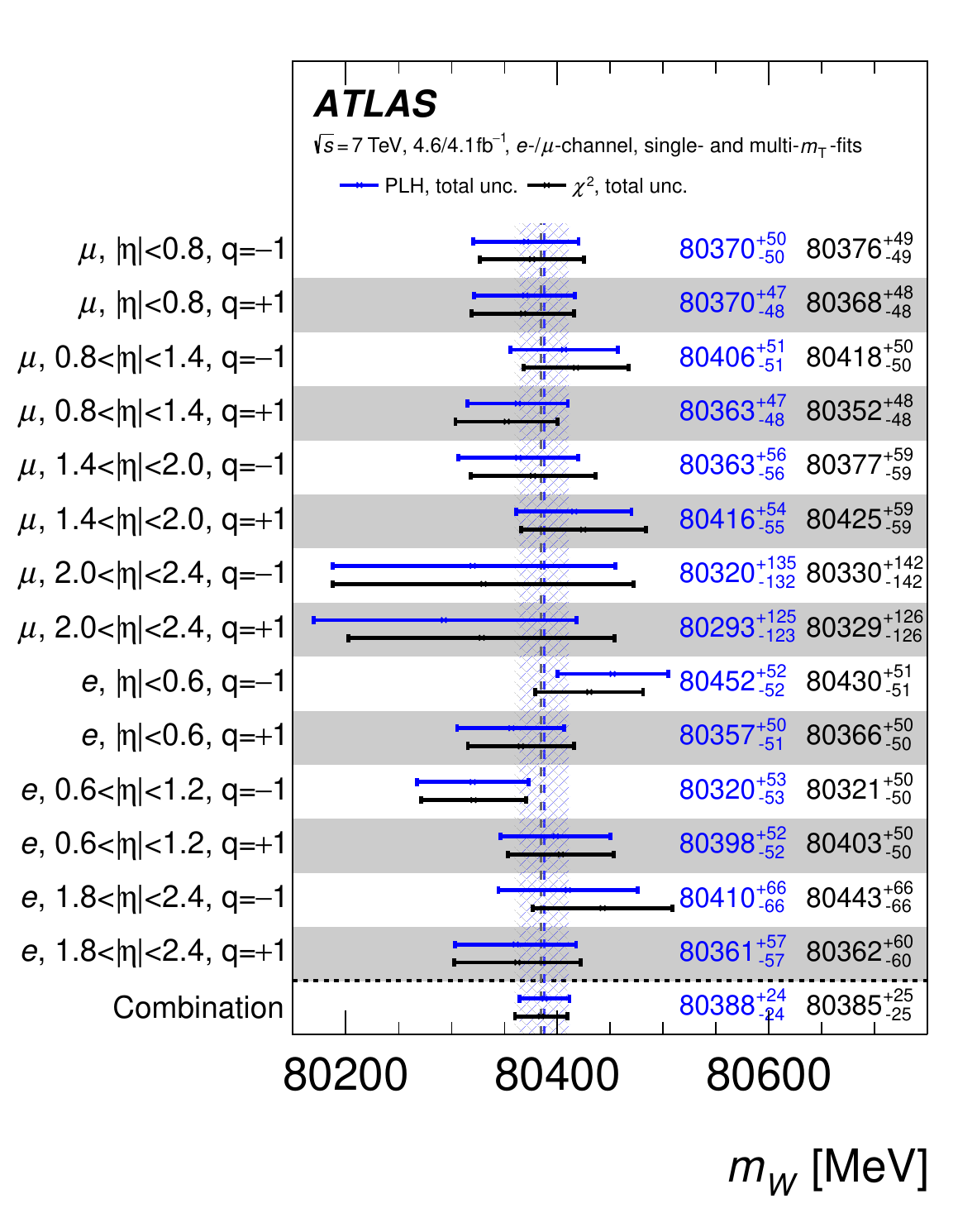}}
\caption{Overview of the \mW fit results in all categories for the (a) \ptl and (b) \mt distributions, with the CT10nnlo PDF set, where $q$ denotes the charge of the decay lepton. The results of the PLH fit are compared with the $\chi^2$ fit, where systematic uncertainties are propagated using the offset method~\cite{STDM-2014-18}. The points labelled as `Combination' correspond to the result of a joint PLH fit to all categories and to a combination of individual $\chi^2$ fits.}
\label{fig:chi2fits}
\end{figure}

The PLH fits using the CT10nnlo PDF set are first performed with statistical uncertainties only. Excellent consistency with the previous results is obtained, which provides a basic validation for the technical aspects of the fit. The comparison is repeated for fits including all systematic uncertainties, with the results summarised in Figure~\ref{fig:chi2fits}. The results of the PLH fits combining all categories yield $\mW = 80357.0 \pm 15.8$~\MeV and $\mW = 80388.2 \pm 23.8$~\MeV for the \ptl\ and \mt\ distributions, respectively. Compared to the original ATLAS measurement, this corresponds to shifts of \mW of $-12.4$~\MeV and $+12.5$~\MeV, respectively, while the total uncertainties are reduced by about 3~\MeV due to the profiling of some systematic uncertainties.
Repeating the present analysis with the fit range used in Ref.~\cite{STDM-2014-18} increases the difference by 3~\MeV.

\begin{figure}[tbp]
\centering
\subfigure[]{\includegraphics[width=0.49\textwidth]{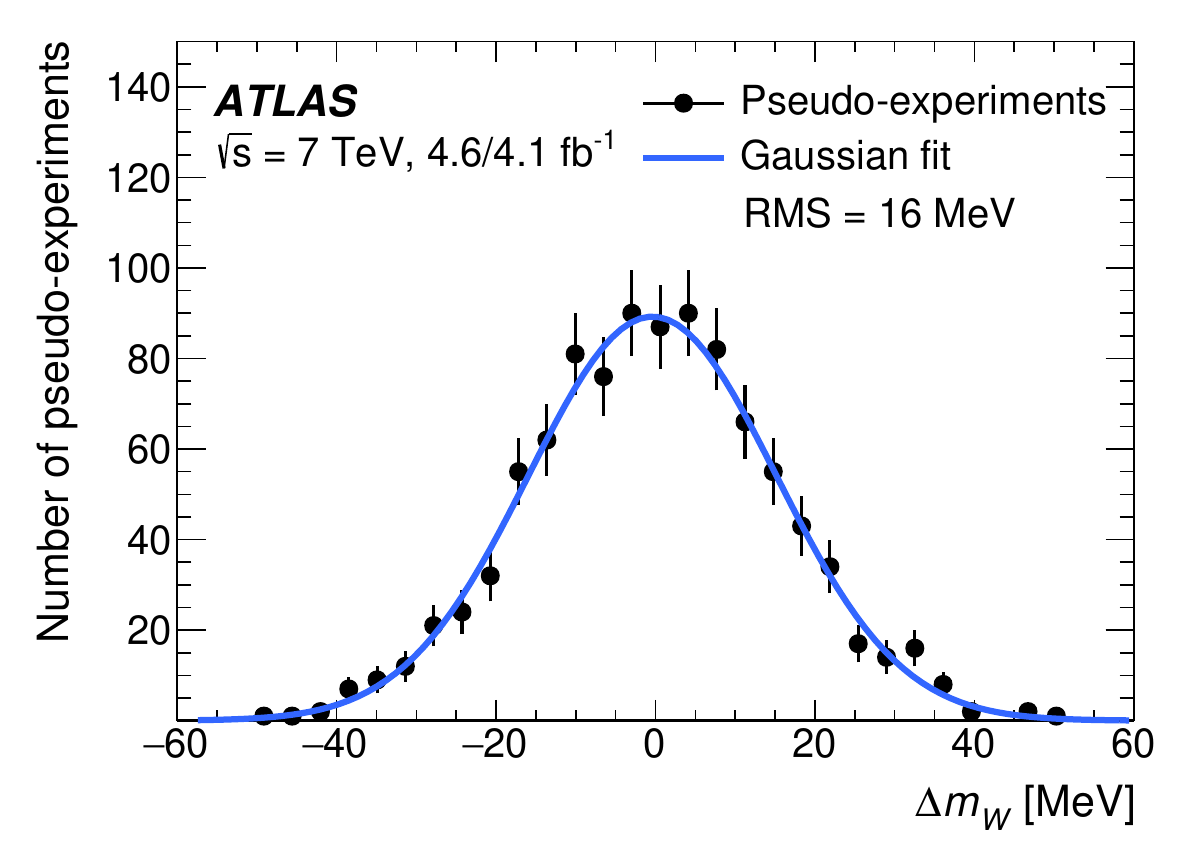}}
\subfigure[]{\includegraphics[width=0.49\textwidth]{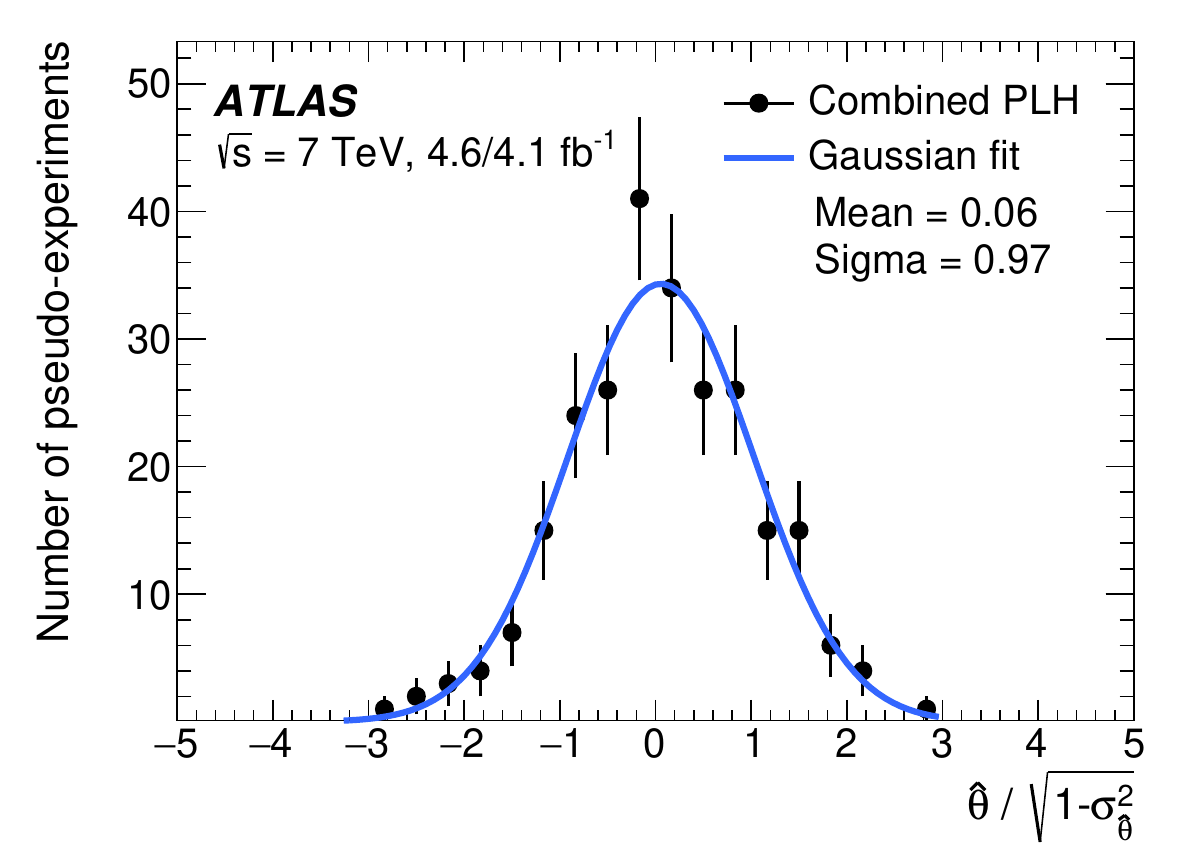}}
\caption{(a) Distribution of the difference $\Delta \mW$ between the nominal \mW PLH fit result and results obtained for pseudo-experiments using random variations of the sources of systematic uncertainty. The \ptl distribution is used. (b) Distribution of pull significances for the NPs in the combined PLH fit to the \ptl distribution.}
\label{fig:toyval-significances}
\end{figure}

The compatibility between the present and previous results is tested by repeating the PLH fits for an ensemble of models where the preferred values of the nuisance parameters are varied randomly within their pre-fit uncertainties. As shown in Figure~\ref{fig:toyval-significances}, the spread of fit results from pseudo-experiments with varied nuisance parameters is about 16~\MeV, confirming that the change in central value introduced by the new statistical method corresponds to about one standard deviation. The distribution of the nuisance parameter pulls\footnote{For a given nuisance parameter $\theta$, the pull is defined as $\hat\theta/\sqrt{1-\sigma_{\hat\theta}}$, where $\hat\theta$ and $\sigma_{\hat\theta}$ are the nuisance parameter post-fit value and uncertainty, respectively.} is consistent with a normal distribution, indicating an overall correct estimate of the pre-fit uncertainties.

\subsection{Impact of updated parton distribution functions}

\begin{figure}[tbp]
\centering
\subfigure[]{\includegraphics[width=0.49\textwidth]{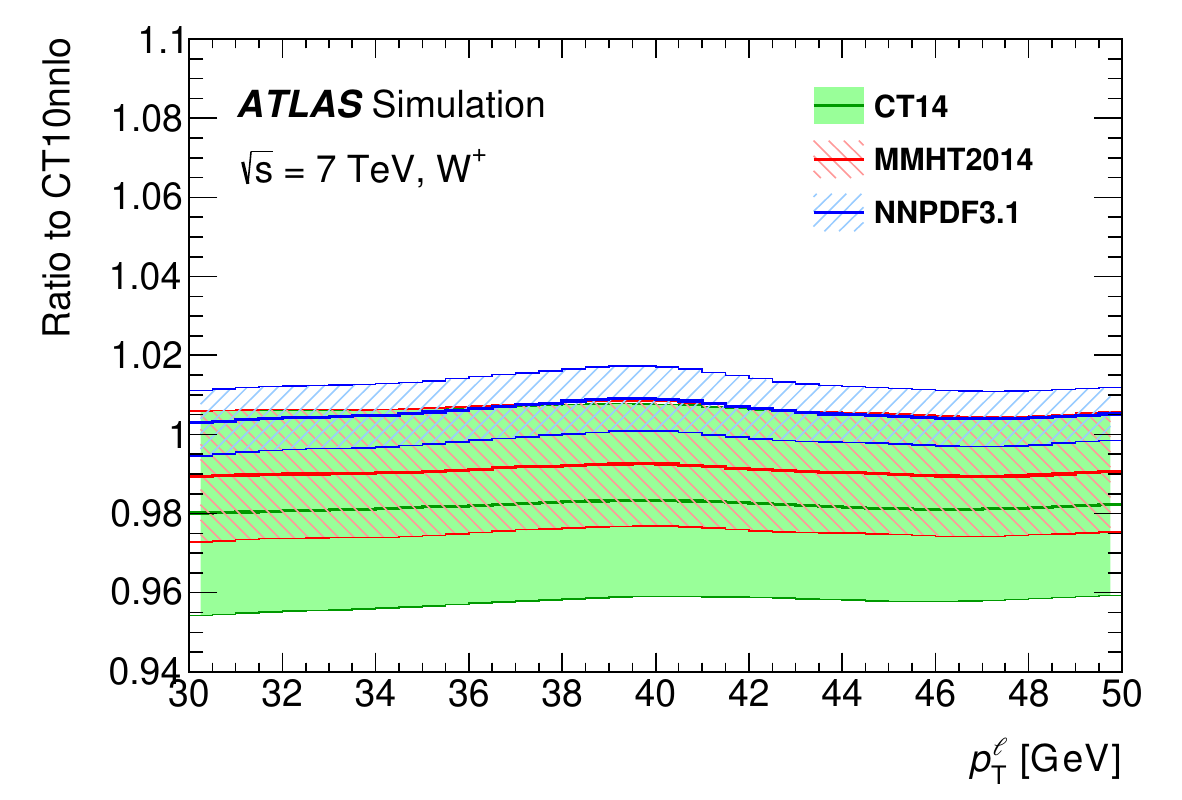}}
\subfigure[]{\includegraphics[width=0.49\textwidth]{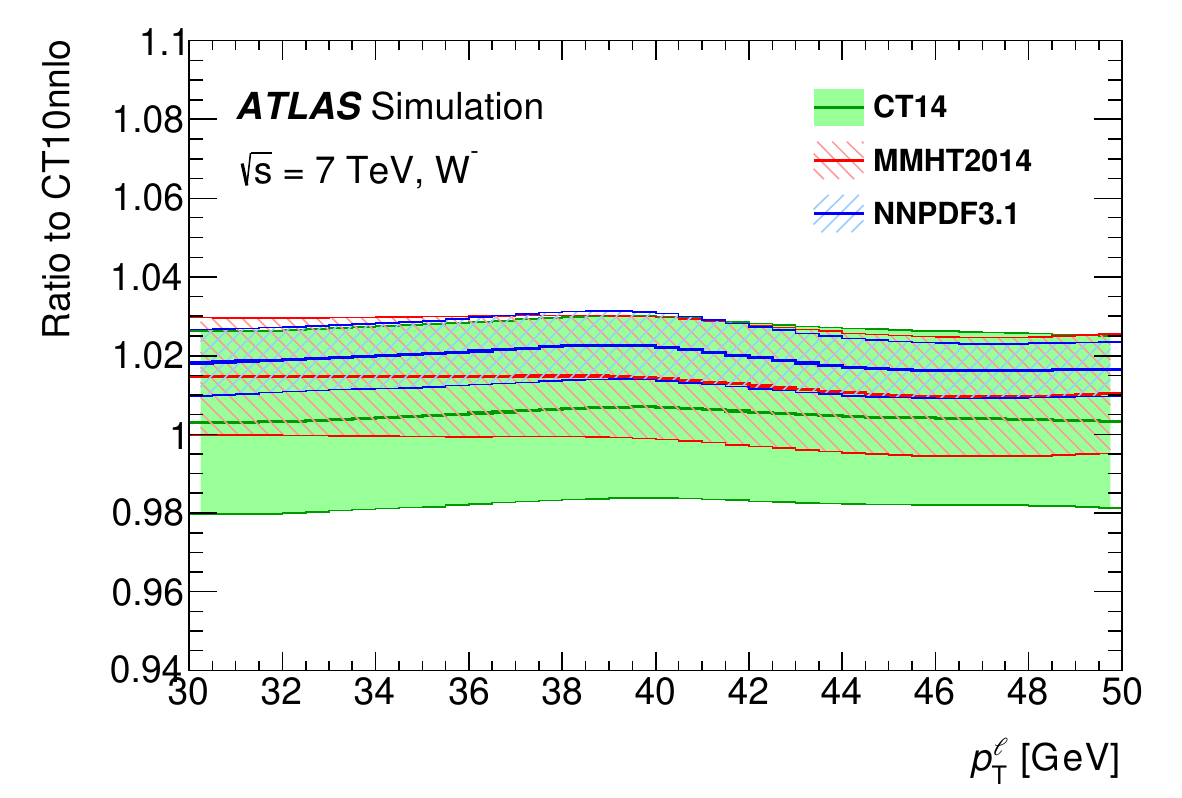}}
\subfigure[]{\includegraphics[width=0.49\textwidth]{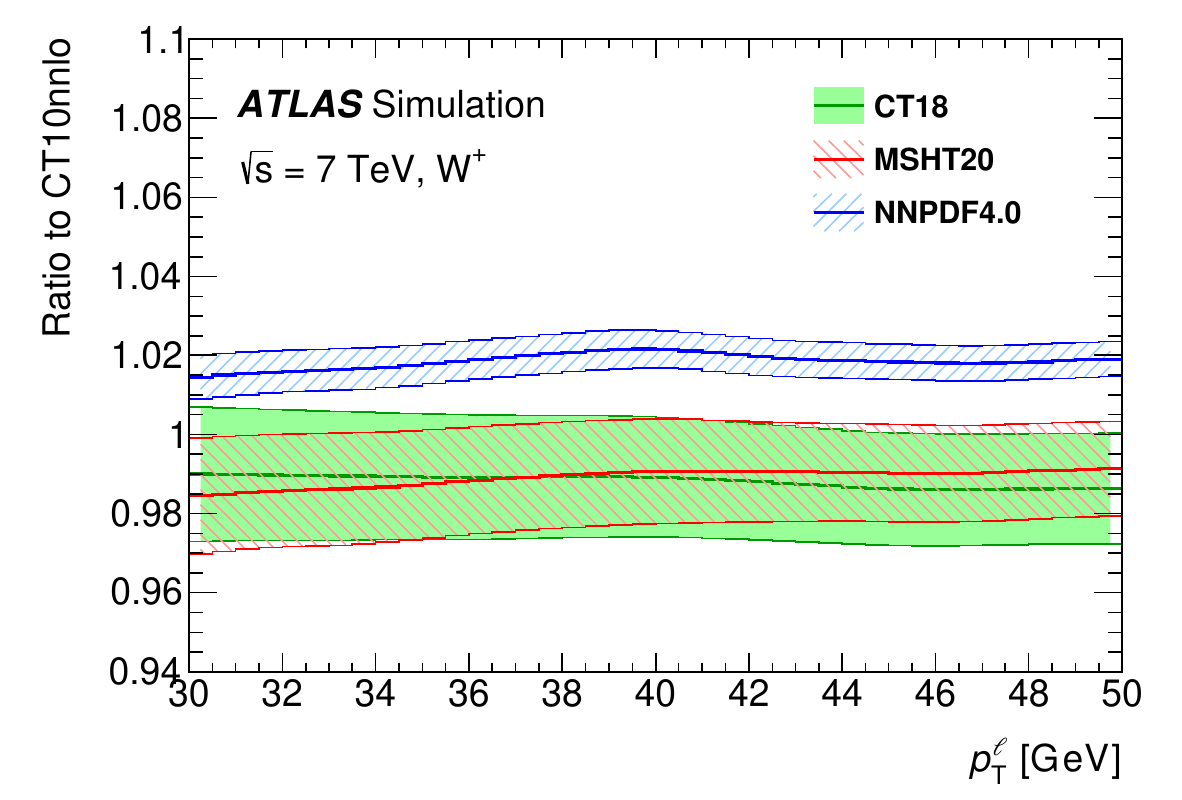}}
\subfigure[]{\includegraphics[width=0.49\textwidth]{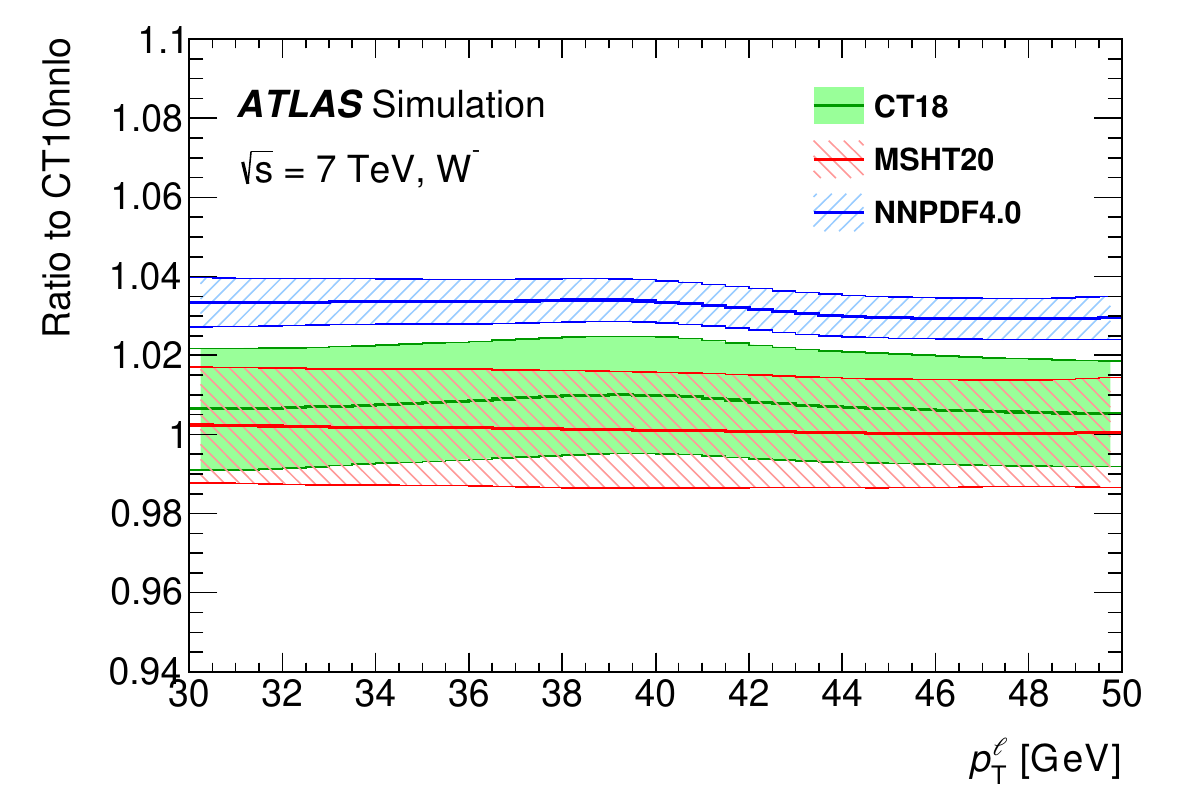}}
\caption{Relative effect, with respect to CT10nnlo, of the indicated PDF extrapolations on the detector-level, $\eta$-inclusive \ptl distributions in (a,c) $W^+$ events and (b,d) $W^-$ events. The thick lines show the central PDF set, and the envelopes show the associated 68\% CL uncertainty.}
\label{fig:pdfextrap}
\end{figure}

The impact of a change in the PDFs on the final state distributions is evaluated using \POWHEG, both to calculate the extrapolation from the central CT10nnlo set to CT14, CT18, CT18A, MMHT2014, MSHT20, NNPDF3.1,NNPDF4.0 and ATLASpdf21, and to calculate the PDF uncertainty variations. All calculations are performed at generator level in full phase space, and the impact on the final-state distributions is evaluated using migration matrices as in Section~\ref{sec:ewuncert}. As in Section~\ref{sec:modunc}, the PDF extrapolations are constrained to leave \ptz unchanged. The impact of the extrapolations on the detector-level \ptl distributions is illustrated in Figure~\ref{fig:pdfextrap}.

\clearpage

\subsection{Results and discussion}

Fit results with updated PDF sets are listed in Table~\ref{tab:fitpdfs}. A satisfactory fit quality is obtained for all PDF sets. Separate fits are performed to the \ptl and \mt distributions as they are projections of the same data, and the corresponding statistical correlations cannot be accounted for in the framework of Eq.~\eqref{eq:plh} in a straightforward way. Moreover, the PCA treatment applied for some classes of systematic uncertainties leads to different sets of nuisance parameters for the two distributions.

\begin{table}[tbp]
\center
\caption{Best-fit value of \mW, total and PDF uncertainties, in~\MeV, and goodness-of-fit for the \ptl and \mt distributions and the PDF sets described in the text. Each fit uses 14 event categories with 40 bins, for 558 degrees of freedom. \label{tab:fitpdfs}}
\begin{tabular}{l|cccc|cccc}
\toprule
& \multicolumn{4}{c|}{\ptl fit} & \multicolumn{4}{c}{\mt fit} \\
PDF set    & \mW & $\sigma_\text{tot}$ & $\sigma_\text{PDF}$ & $\chi^2$/n.d.f. & \mW & $\sigma_\text{tot}$ & $\sigma_\text{PDF}$ & $\chi^2$/n.d.f. \\
\midrule
CT14       & $80358.3$ & $^{+16.1}_{-16.2}$ & 4.6 & 543.3/558 & $80401.3$ & $^{+24.3}_{-24.5}$ & 11.6 & 557.4/558 \\[1mm]
CT18       & $80362.0$ & $^{+16.2}_{-16.2}$ & 4.9 & 529.7/558 & $80394.9$ & $^{+24.3}_{-24.5}$ & 11.7 & 549.2/558 \\[1mm]
CT18A      & $80353.2$ & $^{+15.9}_{-15.8}$ & 4.8 & 525.3/558 & $80384.8$ & $^{+23.5}_{-23.8}$ & 10.9 & 548.4/558 \\[1mm]
MMHT2014   & $80361.6$ & $^{+16.0}_{-16.0}$ & 4.5 & 539.8/558 & $80399.1$ & $^{+23.2}_{-23.5}$ & 10.0 & 561.5/558 \\[1mm]
MSHT20     & $80359.0$ & $^{+13.8}_{-15.4}$ & 4.3 & 550.2/558 & $80391.4$ & $^{+23.6}_{-24.1}$ & 10.0 & 557.3/558 \\[1mm]
ATLASpdf21 & $80362.1$ & $^{+16.9}_{-16.9}$ & 4.2 & 526.9/558 & $80405.5$ & $^{+28.2}_{-27.7}$ & 13.2 & 544.9/558 \\[1mm]
NNPDF3.1    & $80347.5$ & $^{+15.2}_{-15.7}$ & 4.8 & 523.1/558 & $80368.9$ & $^{+22.7}_{-22.9}$ & 9.7 & 556.6/558 \\[1mm]
NNPDF4.0    & $80343.7$ & $^{+15.0}_{-15.0}$ & 4.2 & 539.2/558 & $80363.1$ & $^{+21.4}_{-22.1}$ & 7.7 & 558.8/558 \\[1mm]
\bottomrule
\end{tabular}
\end{table}

The best-fit values of \mW obtained with different PDF sets span a range of about 18~\MeV for the \ptl fits, and about 42~\MeV for the \mt fits. This envelope is dominated by the NNPDF3.1 and NNPDF4.0 fits, which yield the lowest fit values; the range spanned by the other sets is only 9~\MeV for \ptl and 21~\MeV for \mt.

\begin{figure}[tbp]
\centering
\subfigure[]{\includegraphics[width=0.49\textwidth]{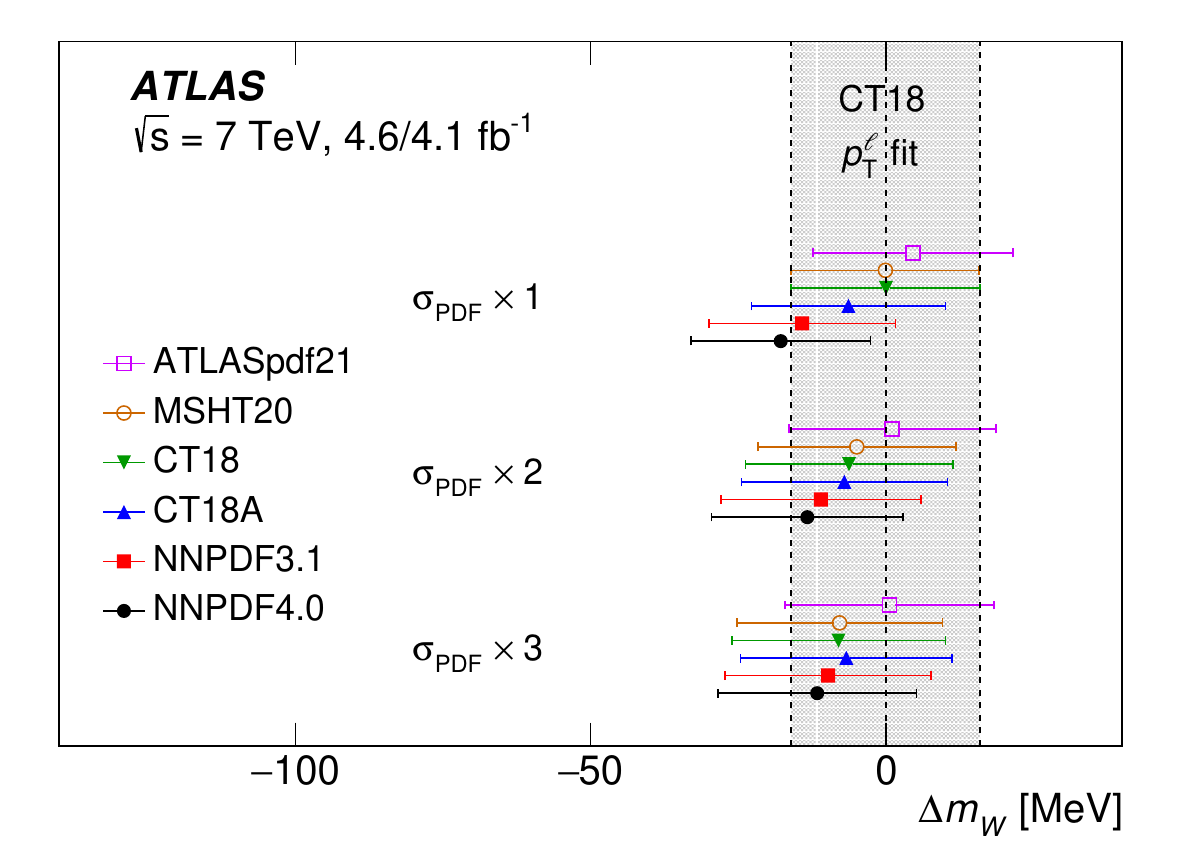}}
\subfigure[]{\includegraphics[width=0.49\textwidth]{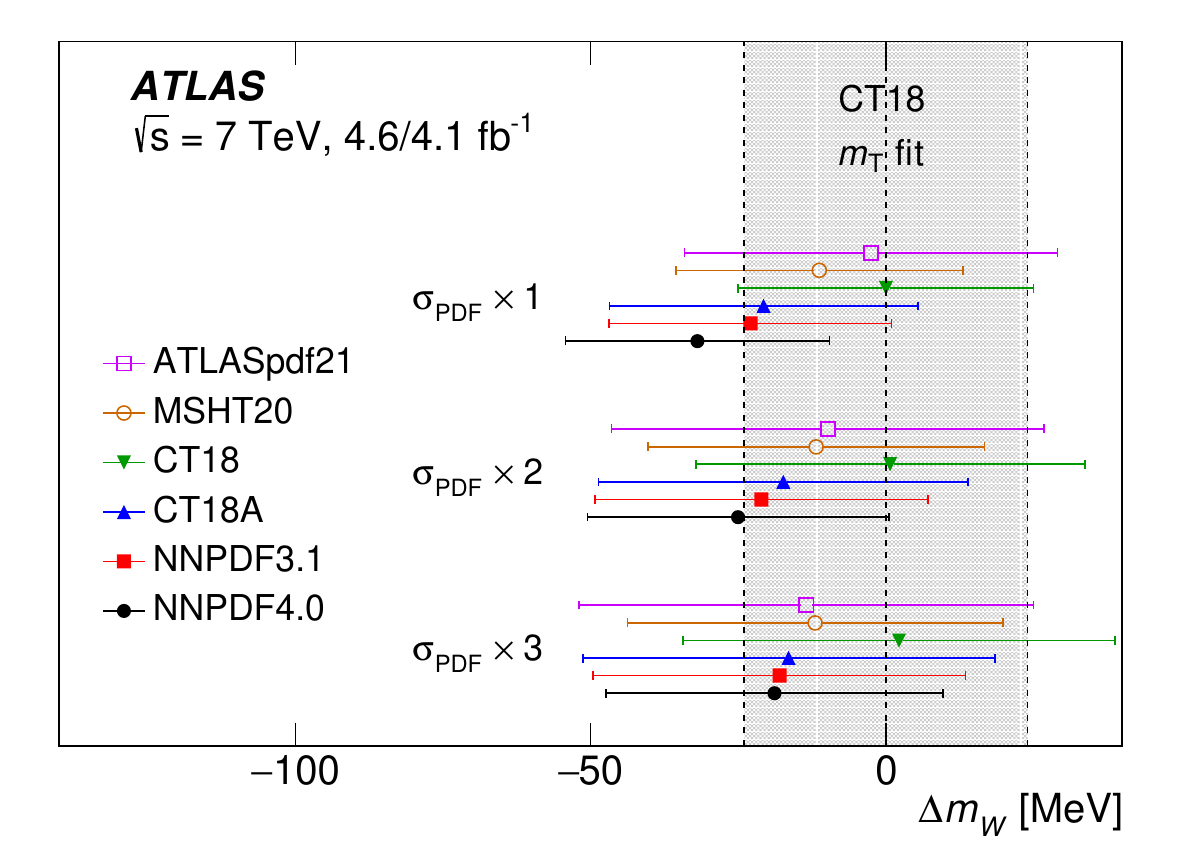}}
\caption{Variation of the fitted value of \mW with the PDF set used in the fit, for the (a)~\ptl and (b)~\mt distributions and different scalings of the pre-fit PDF uncertainties. The reference value is defined by the CT18 PDF set.}
\label{fig:pdferrscale}
\end{figure}

As a cross-check, the influence of the size of the initial PDF uncertainties on the best-fit values is studied in Figure~\ref{fig:pdferrscale}, where the fits are repeated with pre-fit PDF uncertainties scaled by factors 1--3. Enlarged uncertainties allow the models to better adapt to the data, resulting in a reduced PDF model dependence. The differences between the central values found for five out of six PDF sets significantly decrease with larger PDF scaling factors. For factors of 2 and above, the residual model dependence is below 5~\MeV for the \ptl fits, and 25~\MeV for the \mt fits, with the total uncertainty increased by less than 1.5~\MeV.

\begin{figure}[tbp]
\centering
\subfigure[]{\includegraphics[width=0.49\textwidth]{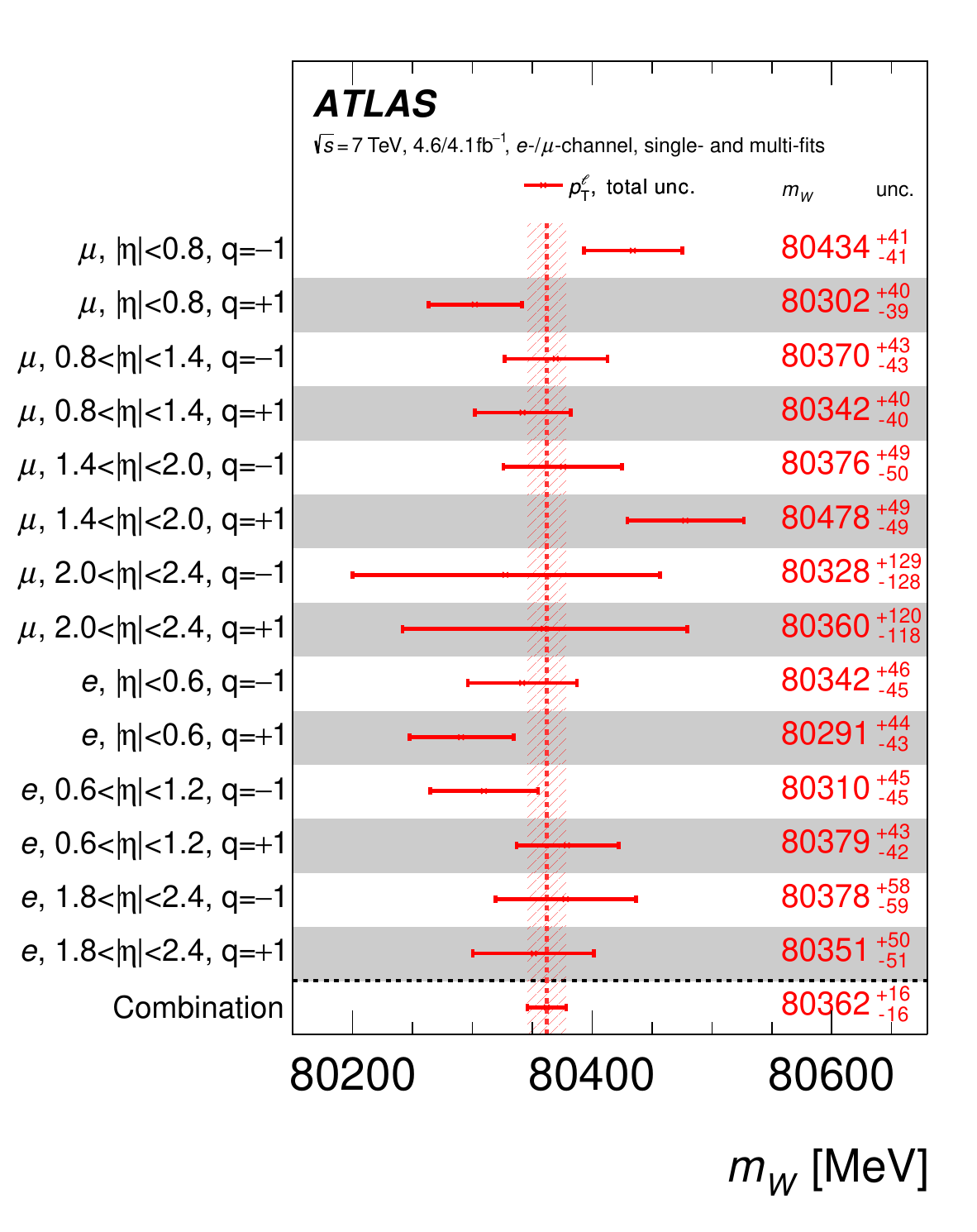}}
\subfigure[]{\includegraphics[width=0.49\textwidth]{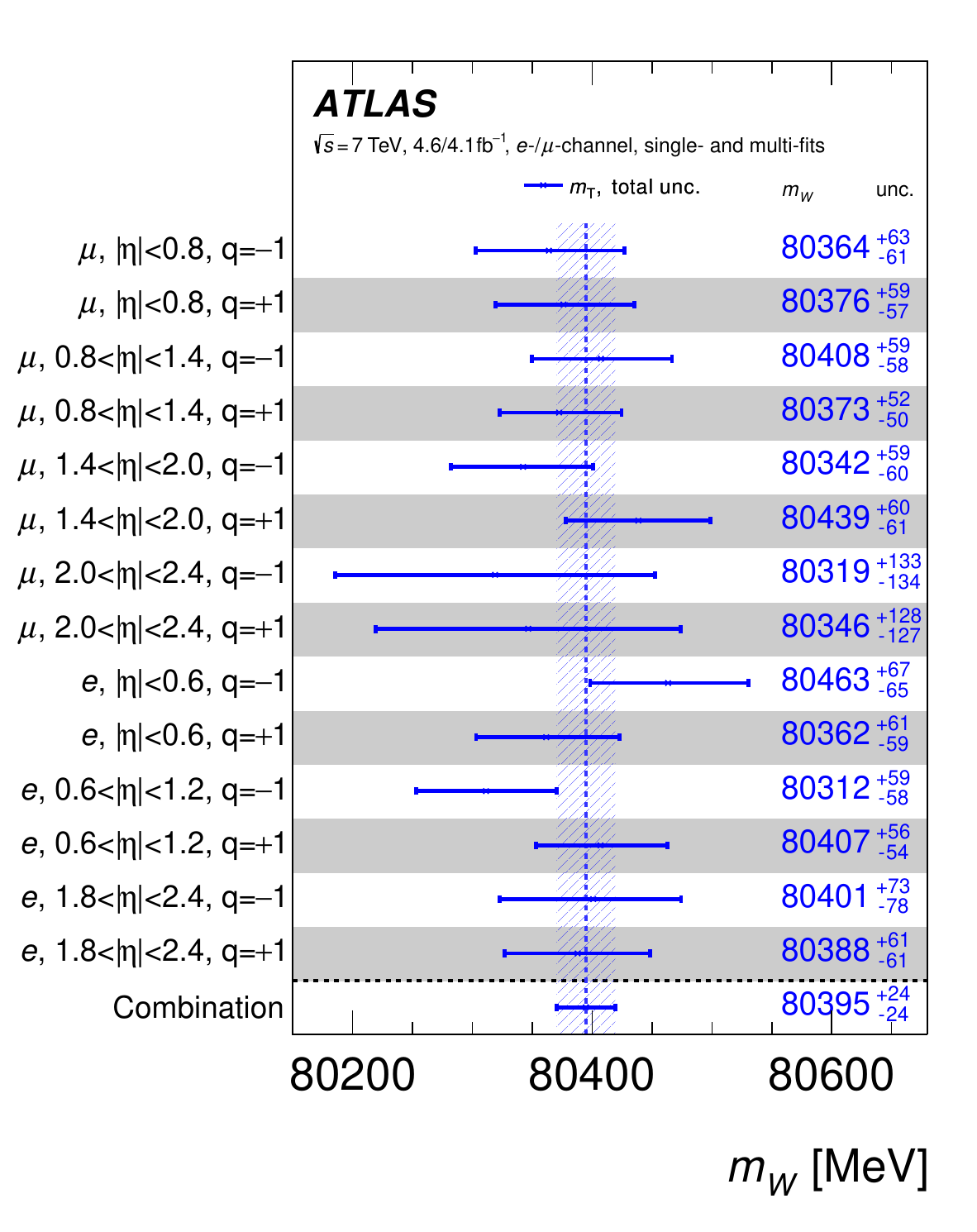}}
\caption{Overview of the \mW PLH fit results in all categories for the (a) \ptl and (b) \mt distributions, with the CT18 PDF set. The points labelled as `Combination' correspond to the result of a joint PLH fit to all categories.}
\label{fig:mwfits}
\end{figure}

\begin{figure}[tbp]
\centering
\subfigure[]{\includegraphics[width=0.49\textwidth]{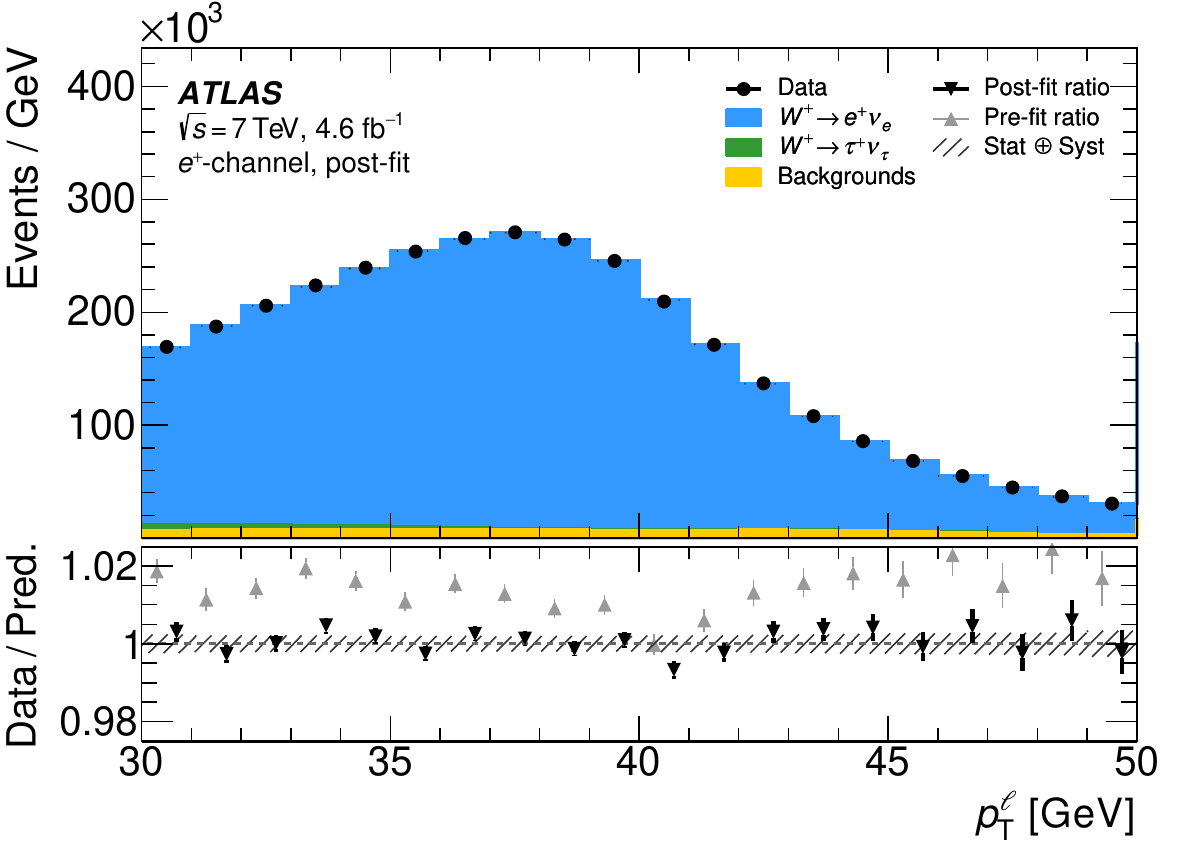}}
\subfigure[]{\includegraphics[width=0.49\textwidth]{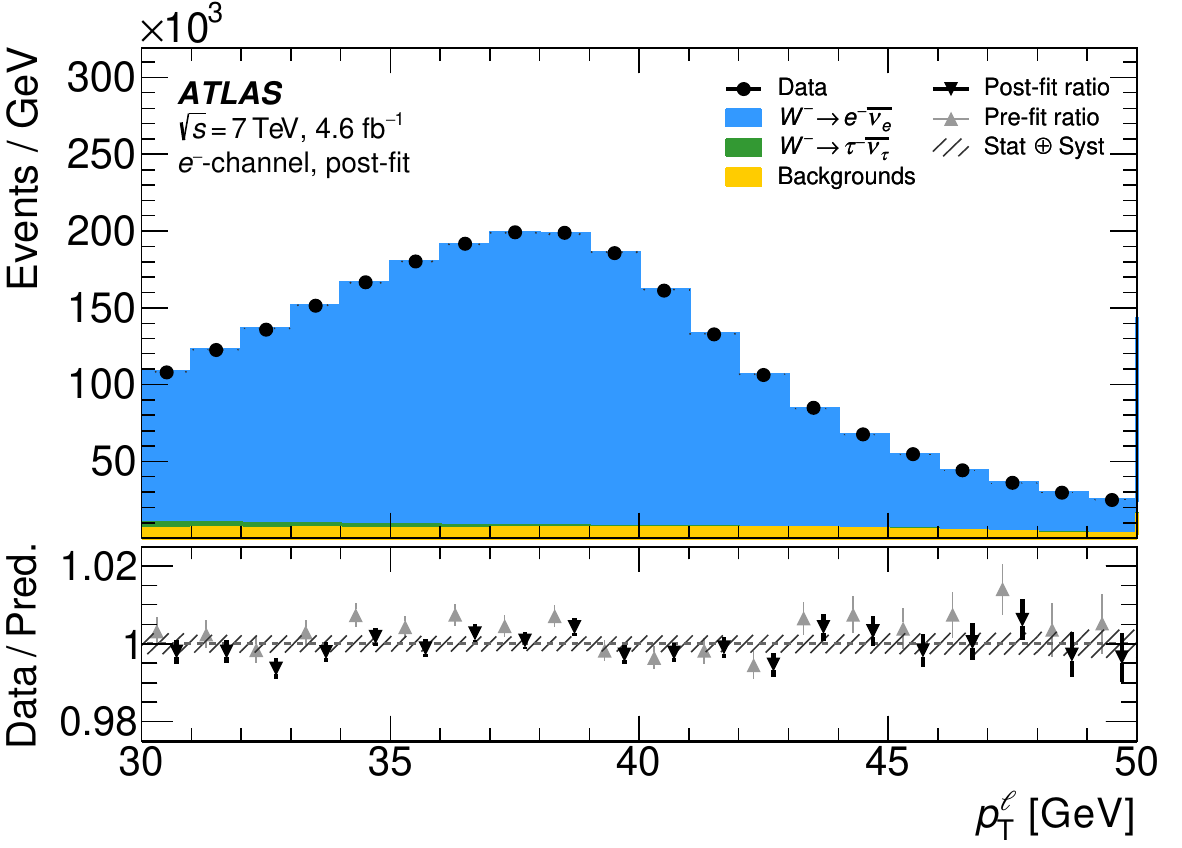}}
\subfigure[]{\includegraphics[width=0.49\textwidth]{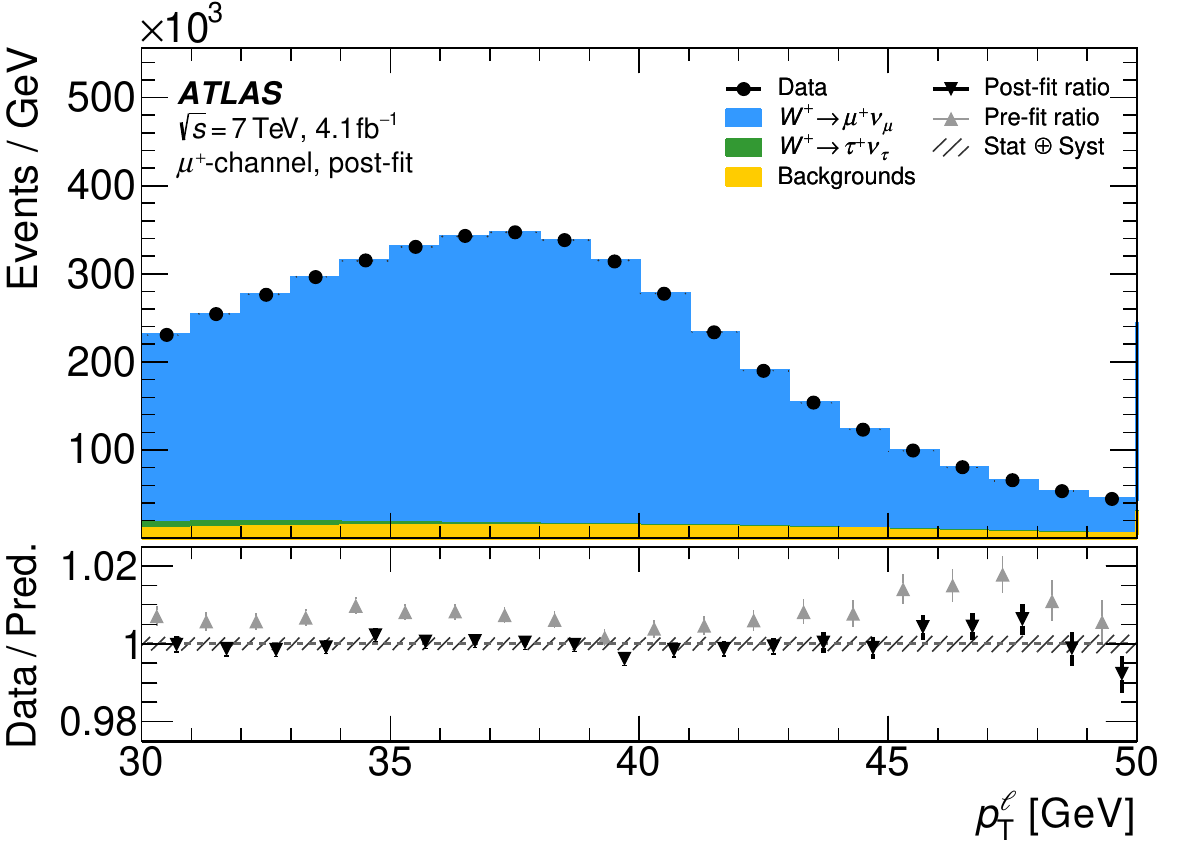}}
\subfigure[]{\includegraphics[width=0.49\textwidth]{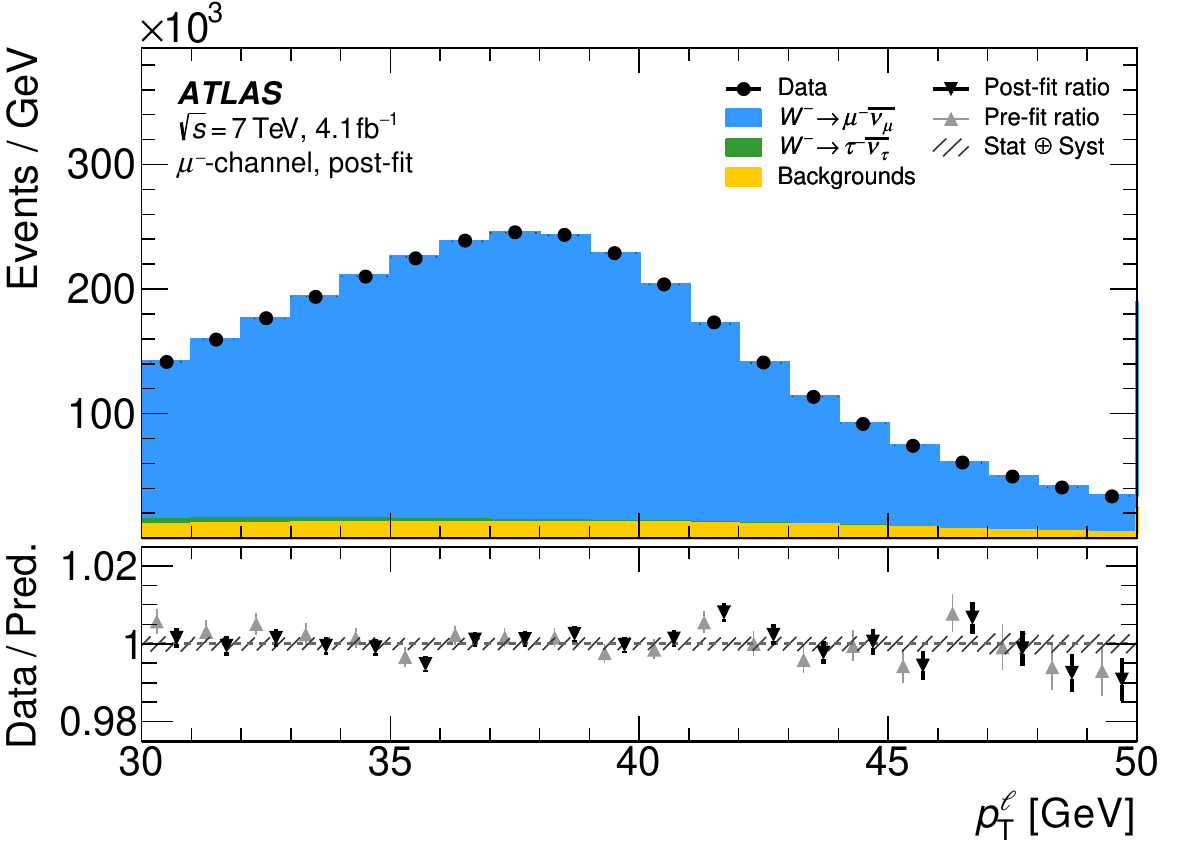}}
\caption{Post-fit distributions of \ptl with data and MC for (a) $W^+\rightarrow e^+\nu_e$, (b) $W^-\rightarrow e^-\overline{\nu}_e$, (c) $W^+\rightarrow \mu^+\nu_{\mu}$ and (d) $W^-\rightarrow \mu^-\overline{\nu}_{\mu}$, inclusive over all $\eta$ regions, and using the CT18 PDF set. In the bottom panels, the darker points represent the post-fit ratio of data to MC, while the lighter points indicate the ratio before the fit. The hatched band represents the total uncertainty of the data.\label{fig:postfit_ptl}}
\end{figure}

The baseline result is defined using CT18 with original uncertainties, which is compatible with the previous exercise and yields the most conservative uncertainty among the PDF sets considered except for ATLASpdf21. CT18 is also the only recent PDF set that does not include the $W$- and $Z$-boson cross sections measured by ATLAS at 7~\TeV~\cite{STDM-2012-20}, which represent the same data as those used in the present analysis. The results for all measurement categories and for the CT18 PDF set are summarised in Figure~\ref{fig:mwfits}. The post-fit, $|\eta|$-inclusive \ptl distributions obtained with CT18 are shown in Figure~\ref{fig:postfit_ptl}, and agree with the data within the uncertainties. Similarly to CT10nnlo, the distribution of the nuisance parameter pulls is consistent with a normal distribution.

\begin{figure}[tbp]
\centering
\subfigure[]{\includegraphics[width=0.49\linewidth]{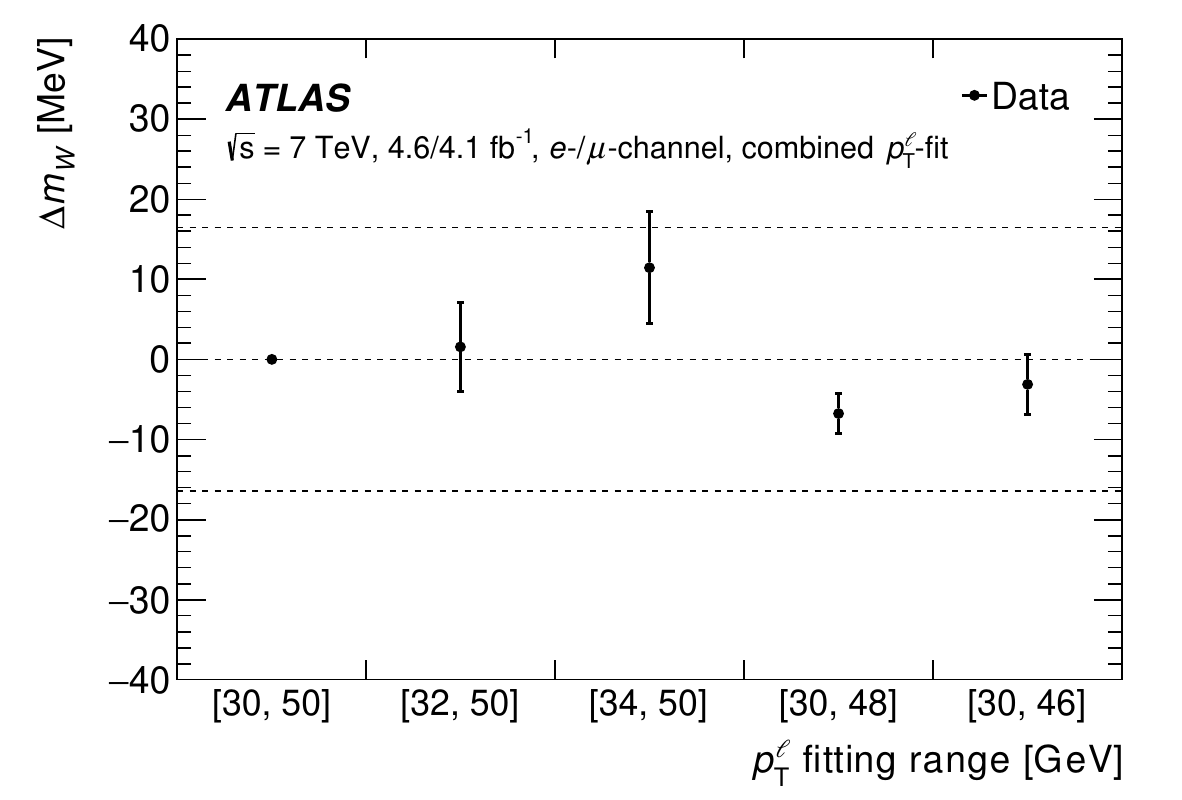}}
\subfigure[]{\includegraphics[width=0.49\linewidth]{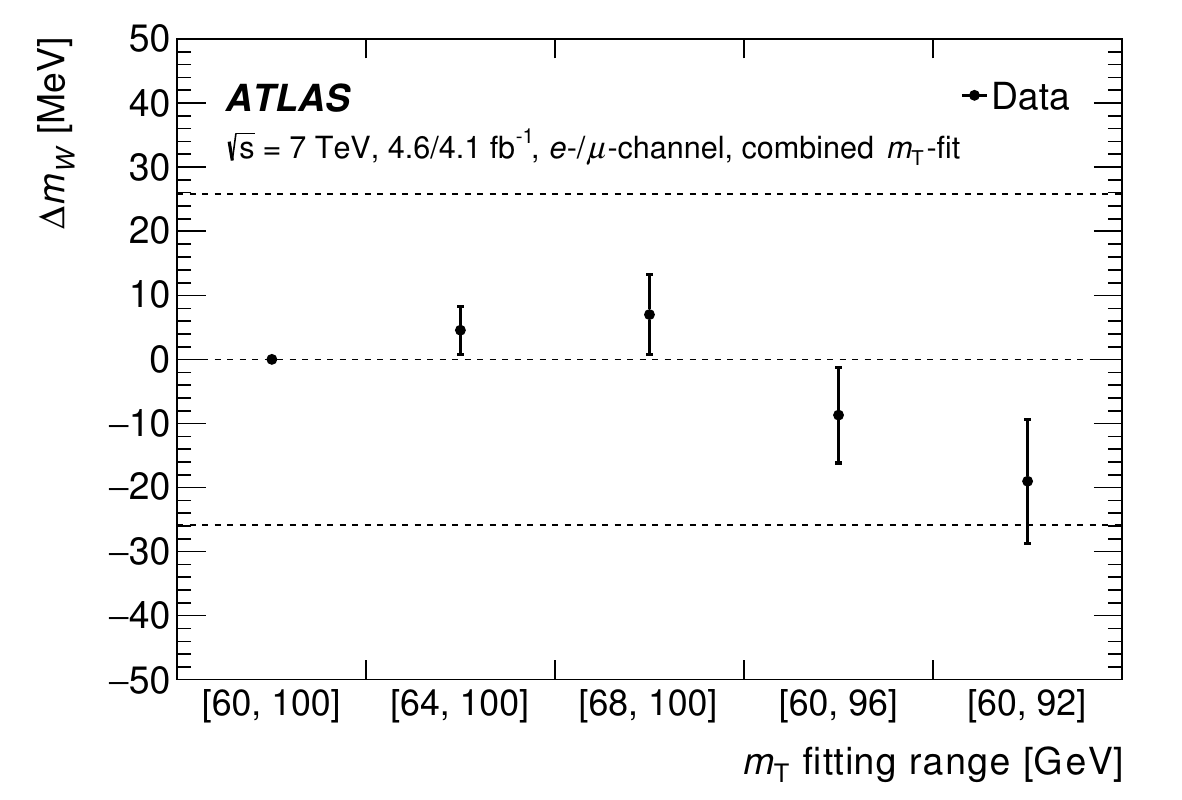}}
\caption{Difference $\Delta \mW$ between the $W$-boson mass measured using the (a) \ptl and (b) \mt distribution fit ranges indicated in the figure and the nominal fit range. The nominal ranges are $30<\ptl<50$~\GeV and $60<\mt<100$~\GeV, respectively. The outer dashed lines indicate the total measurement uncertainty for the nominal range. Results are shown for the combined fit over all categories, and for the CT18 PDF set.\label{fig:wmass:r:rangesCT18}}
\end{figure}

The compatibility of the results for \mW in the different measurement categories is verified by repeating the fit assuming independent parameters of interest in each category. The differences to the baseline fit are small compared with the measurement uncertainties. As a further cross-check, partial fits are performed to the electron and muon channels separately. The electron and muon fit results are found to agree within one standard deviation. A similar exercise is performed for the $W^+$ and $W^-$ channels, and the same conclusion is obtained. Finally, the dependence of the fit result on the \ptl and \mt ranges used for the fit is shown in Figure~\ref{fig:wmass:r:rangesCT18}, with good stability.

\begin{figure}[tbp]
\centering
\subfigure[]{\includegraphics[width=0.49\textwidth]{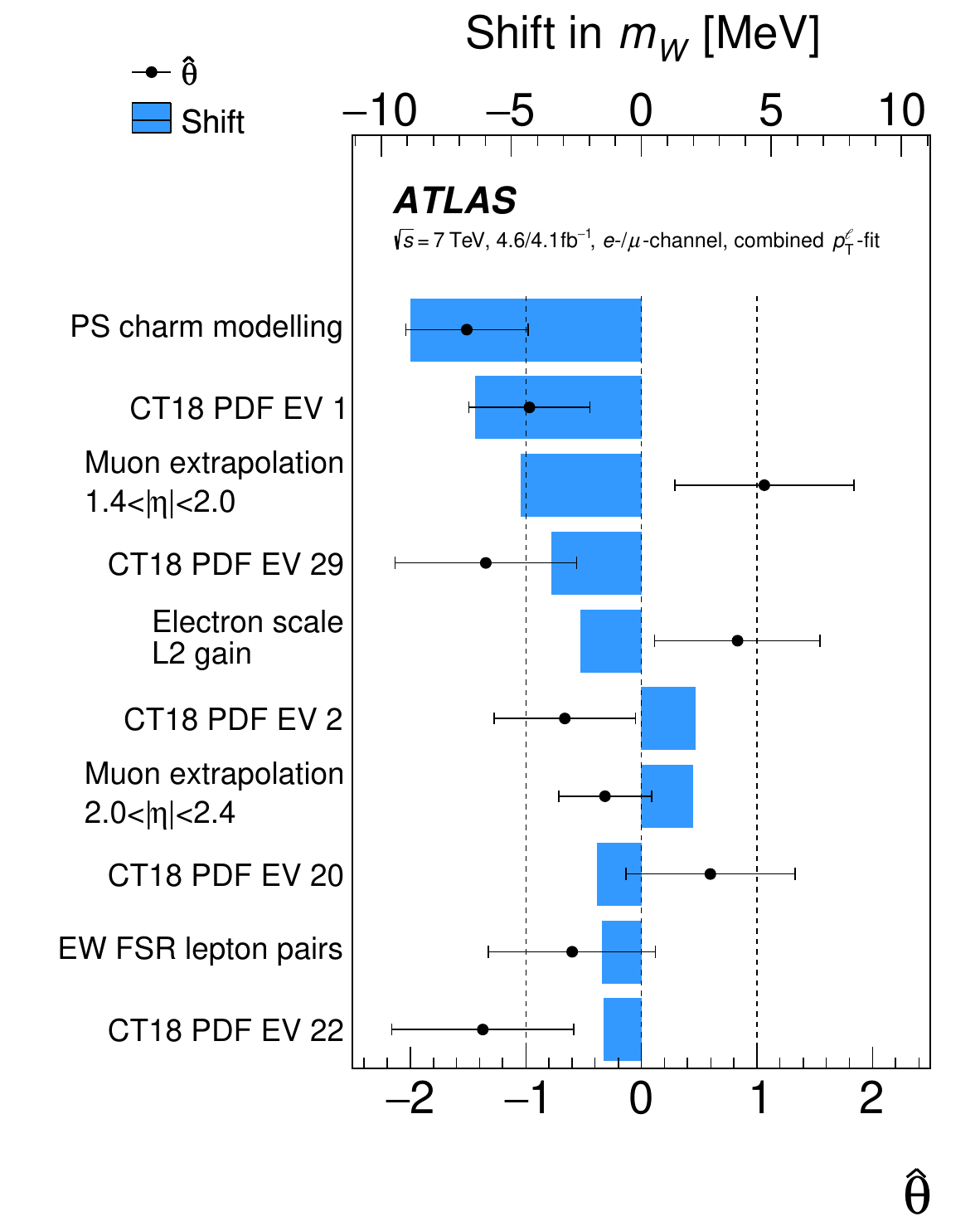}}
\subfigure[]{\includegraphics[width=0.49\textwidth]{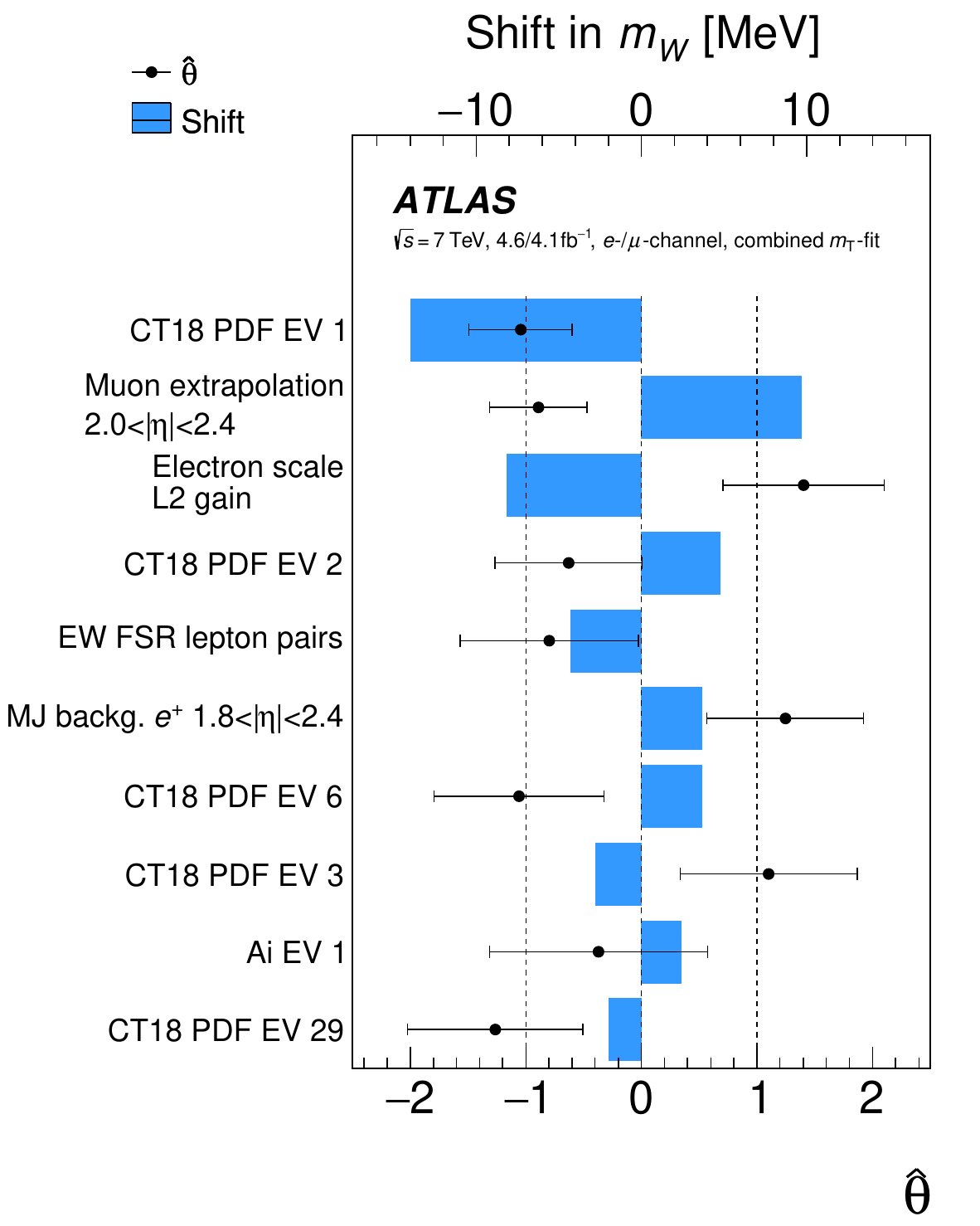}}
\caption{The ten nuisance parameters inducing the largest shifts on the fitted value of \mW in the combined PLH fits, using the (a) \ptl and (b) \mt distributions and the CT18 PDF set. For a given NP $\theta$, the shift is defined as the product of its post-fit value $\hat \theta$ and its pre-fit impact on \mW. The points, which are plotted according to the bottom horizontal scale, show $\hat \theta$ for each of the nuisance parameters. The error bars show the corresponding post-fit uncertainties, $\sigma_{\hat\theta}$. The nuisance parameters are ranked according to the shift induced on \mW, the NPs with the largest shifts at the top.}
\label{fig:mwshiftrank}
\end{figure}

Figure~\ref{fig:mwshiftrank} summarises the ten nuisance parameters that induce the largest shift of \mW in fits to the \ptl and \mt distributions. They are related to electron and muon calibration uncertainties, to the uncertainty in charm-induced production for the \ptw description, to specific eigenvectors~(EV) of the CT18 PDF set, and to missing higher-order electroweak corrections. The corresponding nuisance parameter pulls are also shown. By construction, the PLH fits induce shifts of the nuisance parameters from their nominal value and significant deviations would indicate an underestimation of systematic uncertainties. All observed pulls are within the expectation.

While uncertainties in the PDFs and in the AZ tune parameters have well defined confidence intervals and can be treated as nuisance parameters, this is more questionable for the other uncertainties in the $W$-boson \pt distribution, i.e., the factorisation scale and quark-mass variations. It was verified that the impact of the latter on the final-state distributions is very similar, in shape, to that of the AZ tune parameters, and that these effects are thus not different from the other sources of uncertainty in this respect, and can be treated accordingly. The small post-fit value of the corresponding uncertainty is due to the strong discrimination between the effects of \mW and \ptw variations on the \ptl distribution.

\clearpage

\subsection{Combination\label{sec:mWcomb}}

All event categories are statistically independent as long as only the \ptl or only the \mt distributions are considered. The correlation between the final \ptl- and \mt-based results for \mW is determined from an ensemble of fit results obtained by fluctuating the data and the most probable values of the nuisance parameters within their respective uncertainties. The \ptl and \mt results are then combined using the BLUE prescription~\cite{Lyons:1988rp}. The results of this procedure are given in Table~\ref{tab:combi}. The weight of the \ptl fit ranges from 86\% to 97\%, depending on the PDF set, and dominates the final result. For the CT18 PDF set, the final result is:

$$\mW = 80366.5 \pm 9.8~(\text{stat.}) \pm 12.5~(\text{syst.})~\MeV = 80366.5 \pm 15.9~\MeV,$$

where the first uncertainty component is statistical and the second corresponds to the total systematic uncertainties.

\begin{table}[tbp]
\center
\caption{Uncertainty correlation between the \ptl and \mt fits, combination weights and combination results for \mW and the indicated PDF sets.\label{tab:combi}}
\begin{tabular}{l|cccc}
\toprule
PDF set & Correlation & weight (\ptl) & weight (\mt) & Combined \mW[\MeV] \\
\midrule
CT14       & 52.2\% &  88\% & 12\% & $80363.6 \pm 15.9$ \\
CT18       & 50.4\% &  86\% & 14\% & $80366.5 \pm 15.9$ \\
CT18A      & 53.4\% &  88\% & 12\% & $80357.2 \pm 15.6$ \\
MMHT2014   & 56.0\% &  88\% & 12\% & $80366.2 \pm 15.8$ \\
MSHT20     & 57.6\% &  97\% &  3\% & $80359.3 \pm 14.6$ \\
ATLASpdf21 & 42.8\% &  87\% & 13\% & $80367.6 \pm 16.6$ \\
NNPDF3.1    & 56.8\% &  89\% & 11\% & $80349.6 \pm 15.3$ \\
NNPDF4.0    & 59.5\% &  90\% & 10\% & $80345.6 \pm 14.9$ \\
\bottomrule
\end{tabular}
\end{table}

The decomposition of the post-fit uncertainties is performed according to Ref.~\cite{Pinto:2023yob} and shown in Table~\ref{tab:wmass:r:uncdecomp}. Statistical uncertainties contribute about 10~\MeV in the present fit. This is in contrast with 6~\MeV obtained from fits considering statistical uncertainties only, with all nuisance parameters fixed to their best-fit values. The increase reflects the larger number of parameters determined from the same data. Correspondingly, the systematic uncertainty components are smaller than systematic `impacts' conventionally reported for PLH fits.\footnote{Impacts are obtained from the quadratic subtraction between the total fit uncertainty and the uncertainty of a fit with selected nuisance parameters removed and overestimate the genuine systematic uncertainty.} Systematic uncertainties contribute about 13~\MeV, dominated by PDF uncertainties, missing higher-order electroweak corrections, and electron and muon calibration uncertainties.

\begin{table}[tbp]
\center
\caption{Uncertainty components for the \ptl, \mt and combined \mW measurements using the CT18 PDF set. The first columns give the total, statistical and overall systematic uncertainty in the measurements. The following columns show the contributions of modelling and experimental systematic uncertainties, grouped into categories.\label{tab:wmass:r:uncdecomp}}
\resizebox{\textwidth}{!}{\begin{tabular}{l|ccc|cccccccccc}
\toprule
Unc. [\MeV] & Total  & Stat. & Syst. & PDF & $A_i$ & Backg. & EW & $e$ & $\mu$ & \ut & Lumi & $\GW$ & PS \\
\midrule
\ptl   & 16.2  & 11.1  & 11.8  & 4.9   & 3.5   & 1.7   & 5.6   & 5.9   & 5.4   & 0.9   & 1.1   & 0.1   & 1.5 \\
\mt    & 24.4  & 11.4  & 21.6  & 11.7  & 4.7   & 4.1   & 4.9   & 6.7   & 6.0   & 11.4  & 2.5   & 0.2   & 7.0 \\
Combined & 15.9  & 9.8   & 12.5  & 5.7   & 3.7   & 2.0   & 5.4   & 6.0   & 5.4   & 2.3   & 1.3   & 0.1   & 2.3 \\
\bottomrule
\end{tabular}}
\end{table}%

The fits are performed assuming the SM value for the $W$-boson width, $\GW^\text{SM} = 2088 \pm 1$~\MeV~\cite{deBlas:2021wap}. The fitted value of \mW varies with the assumed value for $\GW$ following $\Delta \mW = -0.06 \, \Delta \GW$. Assuming an alternate SM prediction of $\GW^\text{SM} = 2091 \pm 1$~\MeV, as obtained in Ref.~\cite{Haller:2018nnx}, does not change the measured value of the $W$-boson mass significantly.

\begin{figure}[tbp]
\centering
\subfigure[]{\includegraphics[width=0.49\linewidth]{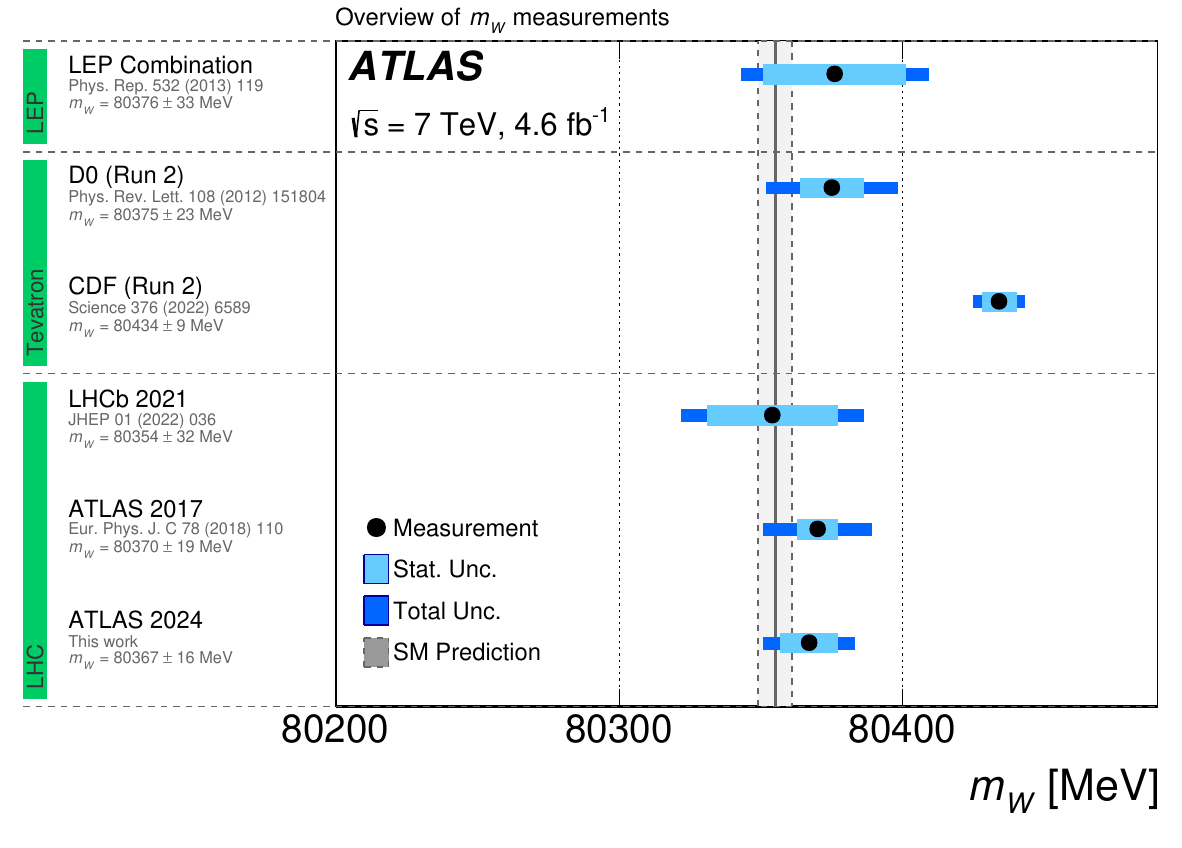}\label{fig:WFinalResults}}
\subfigure[]{\includegraphics[width=0.49\linewidth]{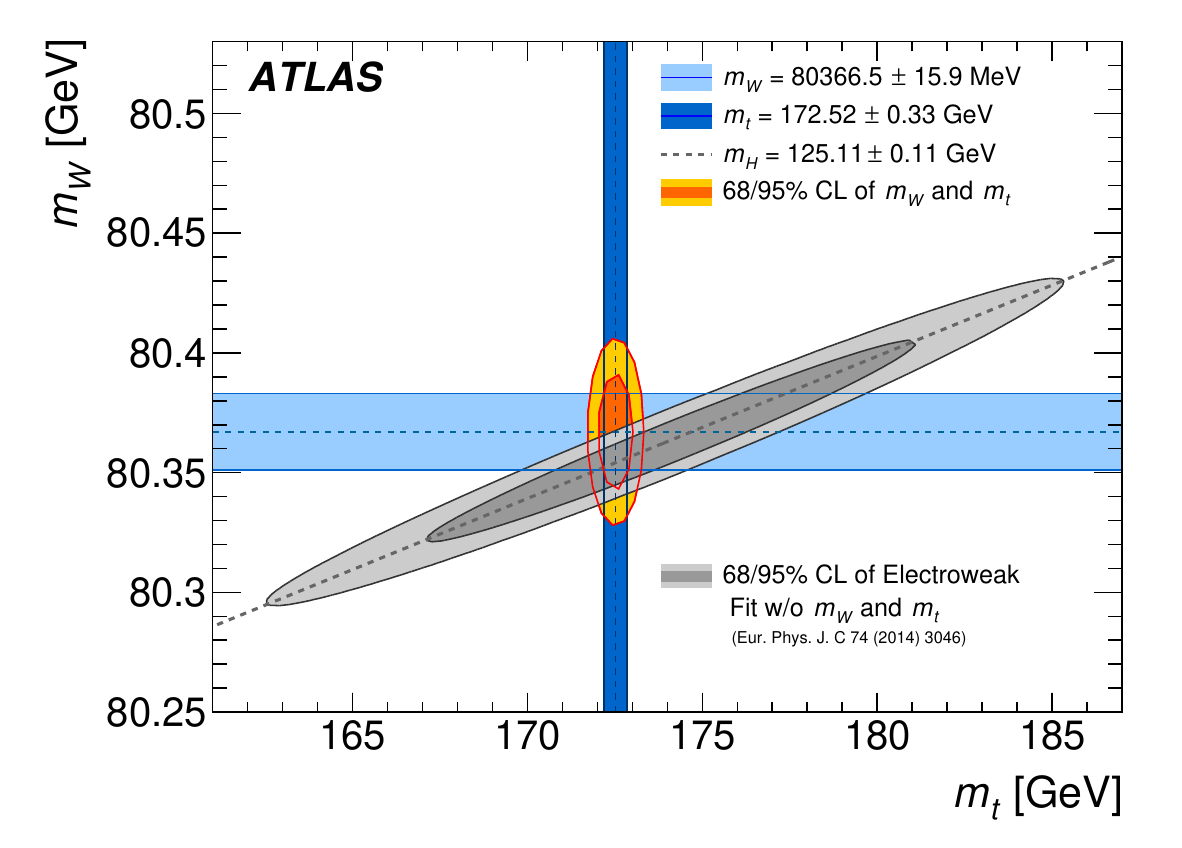}\label{Fig:MWMT}}
\caption{(a) Present measured value of \mW, compared to SM prediction from the global electroweak fit~\cite{deBlas:2021wap}, and to the measurements of LEP~\cite{ALEPH:2013dgf}, Tevatron~\cite{Aaltonen:2012bp, D0:2013jba} and the LHC~\cite{STDM-2014-18, LHCb:2021bjt}. (b) The 68\% and 95\% confidence level contours of the \mW and $m_{t}$ indirect determinations from the global electroweak fit~\cite{Haller:2018nnx}, compared to the 68\% and 95\% confidence-level contours of the present ATLAS measurement of \mW, the ATLAS measurement of $m_H$~\cite{ATLAS:2023oaq} and the LHC measurement of $m_t$~\cite{CMS:2023wnd}.}
\end{figure}

The compatibility of the measured value of the $W$-boson mass using the CT18 PDF set with the Standard Model expectation is illustrated in Figure~\ref{fig:WFinalResults}, together with selected previous measurements. The two-dimensional 68\% and 95\% confidence limits for the predictions of \mW and $m_t$ in the context of the Standard Model electroweak fit are shown in Figure~\ref{Fig:MWMT}, and are compared to the present measurement of \mW and to the combined value of the LHC top-quark mass determinations at 7 and 8~\TeV~\cite{CMS:2023wnd}.

\FloatBarrier


%
\section{Measurement of the $W$-boson width} %

\subsection{Overview}

The \ptl and \mt distributions are not only sensitive to \mW but also to \GW, as shown in Figure~\ref{fig:mwgwtemplates}. In particular, the high tails of the \ptl and \mt distributions are sensitive to changes of \GW. The fit to the \mt distribution is expected to be more sensitive, because events with high \mt are more likely to come from the tail of the $W$-boson Breit--Wigner distribution than events with high \ptl.
The measurement of \GW relies on the same statistical framework, the same calibration, and the same distributions as the previously presented measurement of \mW. However, \GW is left free in the fit, while the $W$-boson mass is treated as NP and set to its SM expectation within the global electroweak fit, $\mW^\text{SM} = 80355 \pm 6$~\MeV \cite{deBlas:2021wap}.
The templates are generated with different values of \GW, centred around the reference value used in the Monte Carlo signal samples. All results are obtained using the same fit ranges as in the \mW measurement: $60 < \mt < 100$~\GeV and $30 < \ptl < 50$~\GeV. The choice of fitting range is driven by the uncertainties in the lepton performance and the hadronic recoil.

\subsection{Results and discussion}
\label{sec:gWresults}

\begin{figure}[tbp]
\centering
\subfigure[]{\includegraphics[width=0.49\textwidth]{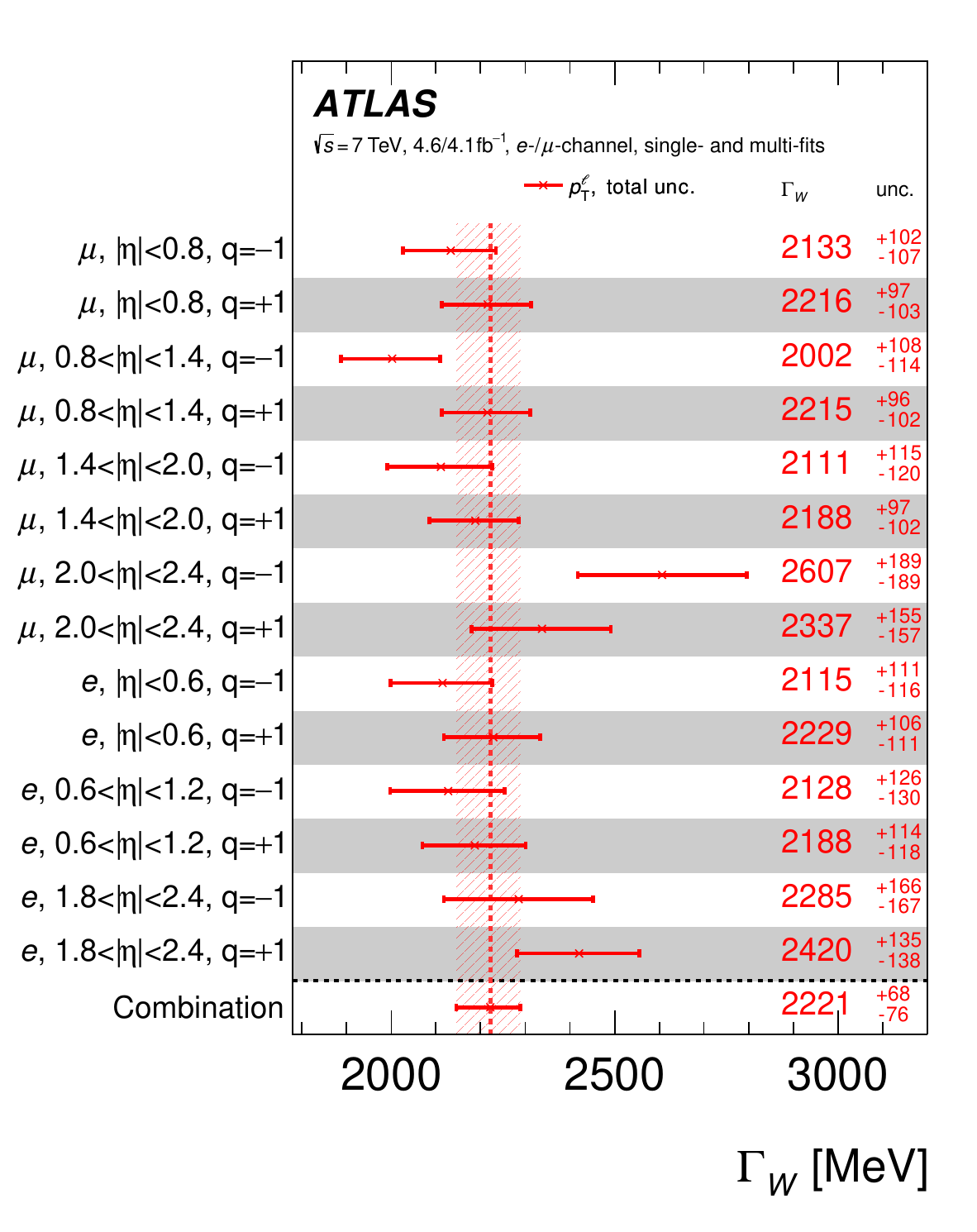}}
\subfigure[]{\includegraphics[width=0.49\textwidth]{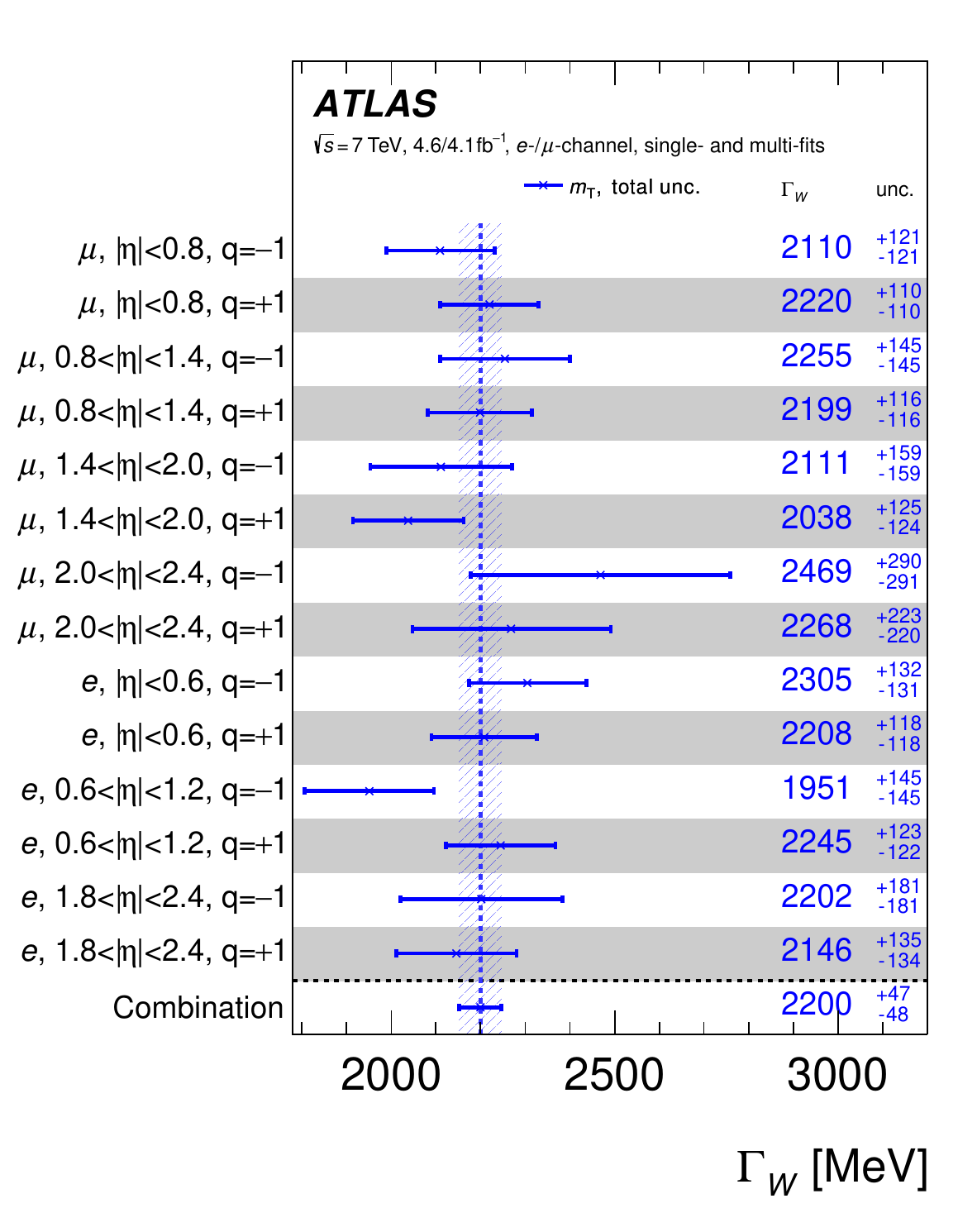}}
\caption{Overview of the \GW PLH fit results in all categories for the (a) \ptl and (b) \mt distributions, with the CT18 PDF set. The points labelled as `Combination' correspond to the result of a joint PLH fit to all categories.}
\label{fig:gwfits}
\end{figure}

The results for each measurement category including all systematic uncertainties for the CT18 PDF set are summarised in Figure~\ref{fig:gwfits} yielding the values of $\GW = 2221^{+68}_{-76}$~\MeV and $\GW = 2200^{+47}_{-48}$~\MeV for \ptl and \mt distributions respectively. Good agreement between the categories can be observed.

Contrary to \mW, the fitted value of \GW depends more strongly on the assumed value of the mass.
The fitted value of \GW varies with the assumed value for \mW following $\Delta \GW = -1.25 \, \Delta \mW$.
In the Standard Model, the predicted value of \mW mainly depends on the assumed value of $m_t$. The present result is based on Ref.~\cite{deBlas:2021wap}, which uses $m_t=172.6$~\GeV and is close to the LHC combined value used in Figure~\ref{Fig:MWMT}. Using $m_t=171.8$~\GeV~\cite{deBlas:2022hdk} or $m_t=173.1$~\GeV~\cite{deBlas:2017wmn} yields $\mW^\text{SM}=80350$~\MeV or 80360~\MeV, respectively, with corresponding variations of \GW by $\pm 6$~\MeV.

The impact of different PDF sets (CT14, CT18, CT18A, MMHT2014, MSHT20, ATLASpdf21, NNPDF3.1, NNPDF4.0) on the \GW measurement is also studied. The results of full fits for all considered PDF sets are summarised in Table~\ref{tab:tabGWPDFs}, with again a satisfactory fit quality for all PDF sets.
The PDF dependence of the fit result is weaker than for \mW, and all central values are well within the uncertainties obtained with CT18. The CT18 PDF set is chosen for the baseline result, consistently with the \mW measurement.

\begin{table}[tbp]
\center
\caption{Best-fit value of \GW, total and PDF uncertainties, in~\MeV, and goodness-of-fit for the \ptl and \mt distributions and the PDF sets described in the text. Each fit uses 14 event categories with 40 bins, for 558 degrees of freedom. \label{tab:tabGWPDFs}}
\begin{tabular}{l|cccc|cccc}
\toprule
& \multicolumn{4}{c|}{\ptl fit} & \multicolumn{4}{c}{\mt fit} \\
PDF set    & $\GW$ & $\sigma_\text{tot}$ & $\sigma_\text{PDF}$ & $\chi^2$/n.d.f. & $\GW$ & $\sigma_\text{tot}$ & $\sigma_\text{PDF}$ & $\chi^2$/n.d.f. \\
\midrule
CT14       & $2228$ & $^{+67}_{-83}$ & 24 & 550.0/558 & $2202$ & $^{+48}_{-48}$ & 5 & 556.8/558 \\[1mm]
CT18       & $2221$ & $^{+68}_{-76}$ & 21 & 534.5/558 & $2200$ & $^{+47}_{-48}$ & 5 & 548.8/558 \\[1mm]
CT18A      & $2207$ & $^{+68}_{-75}$ & 18 & 533.0/558 & $2181$ & $^{+47}_{-48}$ & 5 & 550.6/558 \\[1mm]
MMHT2014   & $2155$ & $^{+71}_{-78}$ & 19 & 546.0/558 & $2186$ & $^{+48}_{-48}$ & 5 & 562.2/558 \\[1mm]
MSHT20     & $2206$ & $^{+66}_{-79}$ & 15 & 556.5/558 & $2179$ & $^{+47}_{-48}$ & 4 & 559.4/558 \\[1mm]
ATLASpdf21 & $2213$ & $^{+67}_{-73}$ & 18 & 531.3/558 & $2190$ & $^{+47}_{-48}$ & 6 & 545.6/558 \\[1mm]
NNPDF31    & $2203$ & $^{+65}_{-78}$ & 20 & 531.7/558 & $2180$ & $^{+47}_{-47}$ & 6 & 560.4/558 \\[1mm]
NNPDF40    & $2182$ & $^{+69}_{-68}$ & 12 & 550.5/558 & $2184$ & $^{+47}_{-47}$ & 4 & 564.0/558 \\[1mm]
\bottomrule
\end{tabular}
\end{table}

\begin{figure}[tbp]
\centering
\subfigure[]{\includegraphics[width=0.49\linewidth]{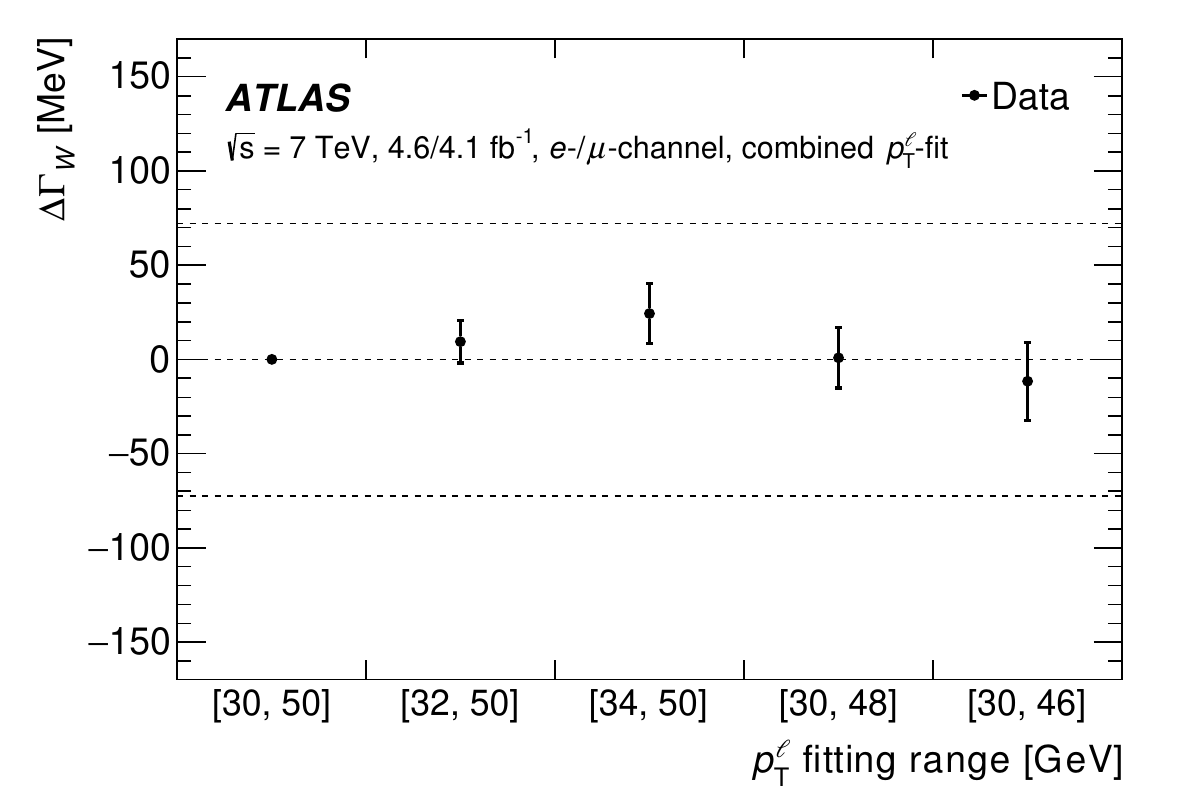}}
\subfigure[]{\includegraphics[width=0.49\linewidth]{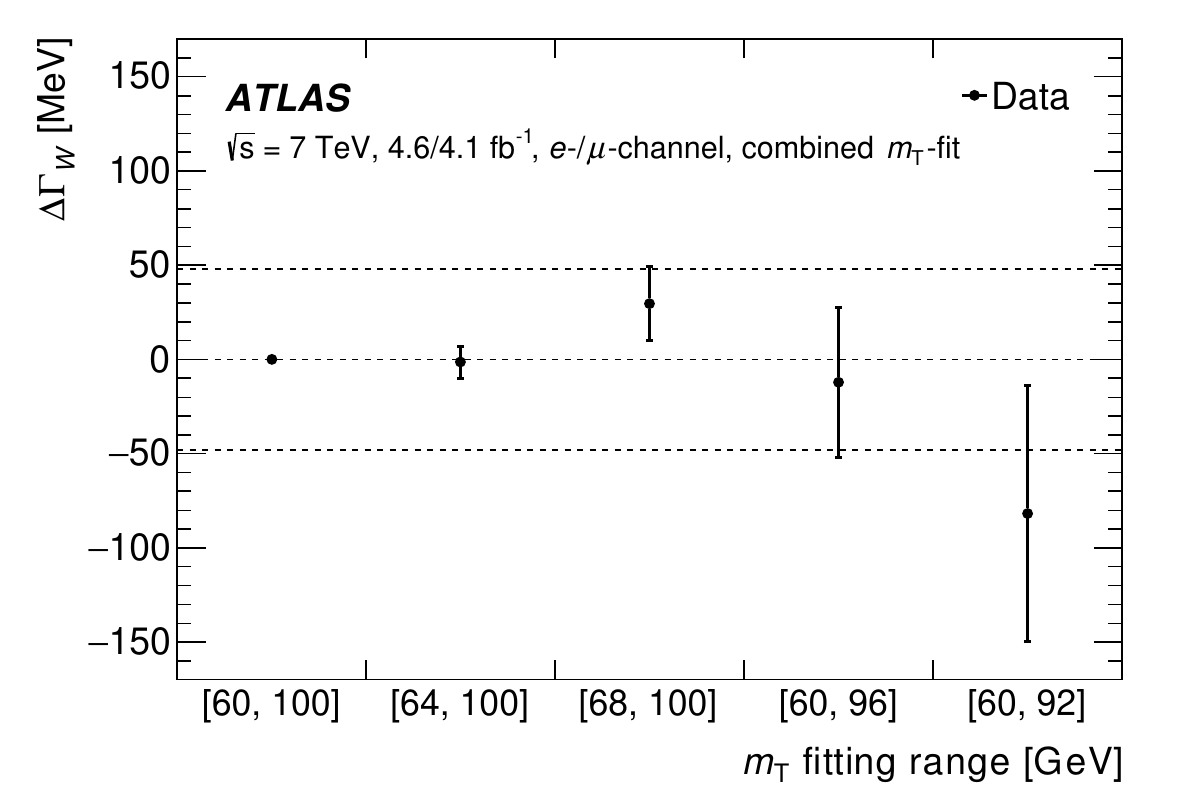}}
\caption{Difference $\Delta \GW$ between the $W$-boson width measured using the (a) \ptl and (b) \mt distribution fit ranges indicated in the figure and the nominal fit range. The nominal ranges are $30<\ptl<50$~\GeV and $60<\mt<100$~\GeV, respectively. The outer dashed lines indicate the total measurement uncertainty for the nominal range. Results are shown for the combined fit over all categories, and for the CT18 PDF set.\label{fig:gwrangesCT18}}
\end{figure}

As a check of compatibility, partial fits are performed to the electron and muon channels separately. These fit results are found to agree within one standard deviation. Similarly, separate fits are performed in the $W^+$ and $W^-$ channels. The results for the two charges are consistent within the 68\%~CL contour of the two-dimensional likelihood function for the \mt fits, while for the \ptl fits the consistency between the two charges is within two standard deviations. Finally, the dependence of the measurement result on the \mt and \ptl ranges used for the fit is studied in Figure~\ref{fig:gwrangesCT18}, with stable results.

\begin{figure}[tbp]
\centering
\subfigure[]{\includegraphics[width=0.49\textwidth]{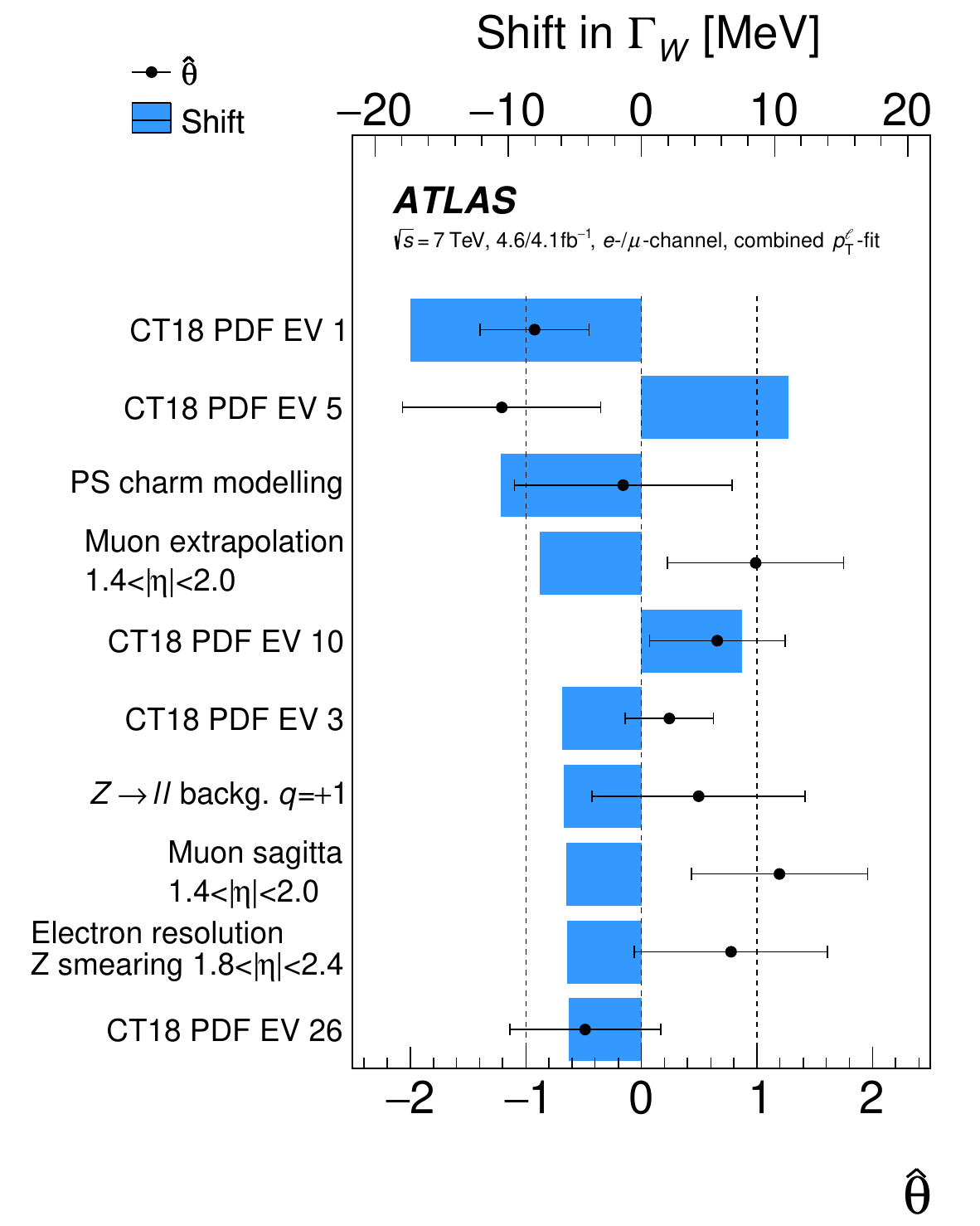}}
\subfigure[]{\includegraphics[width=0.49\textwidth]{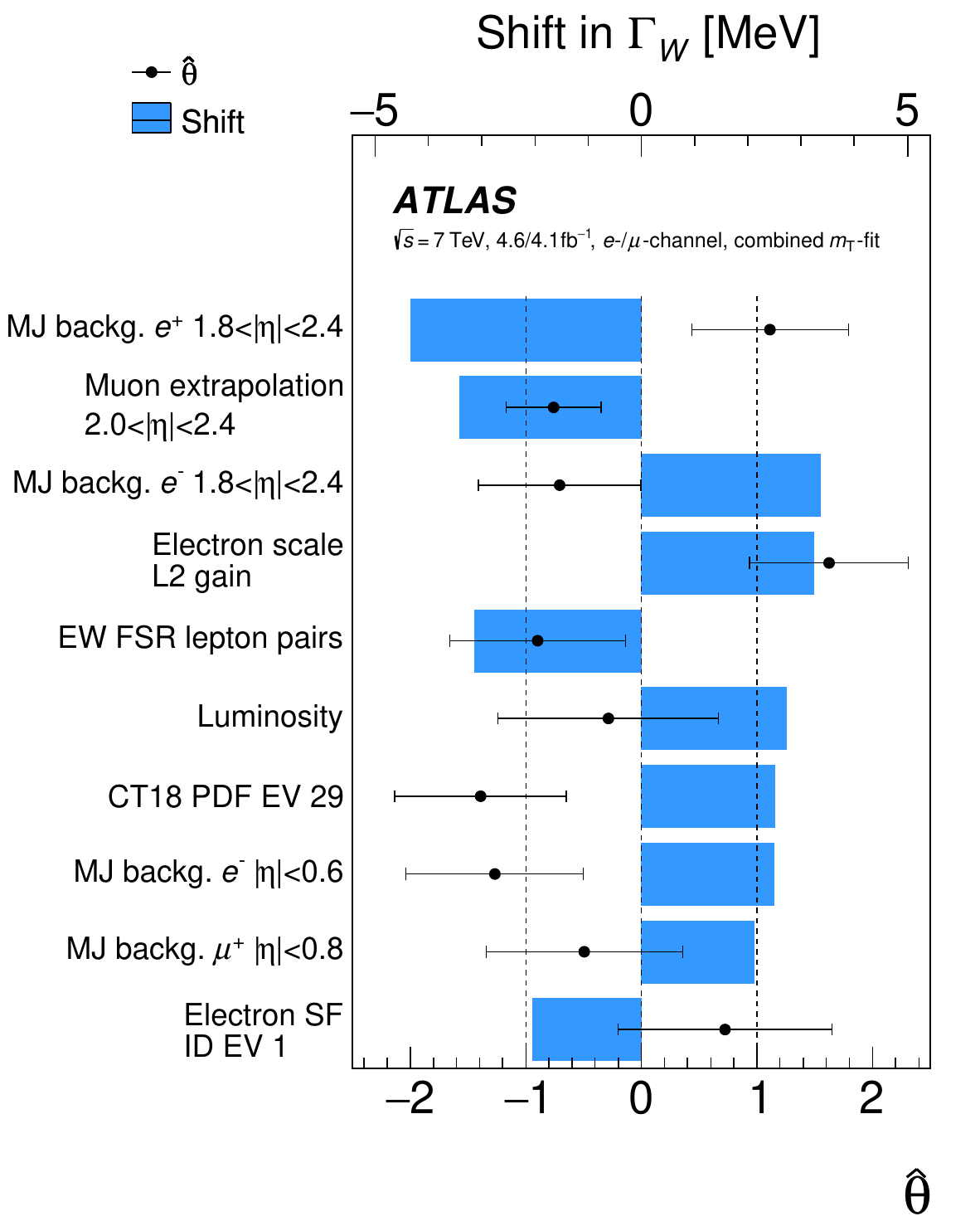}}
\caption{The ten nuisance parameters inducing the largest shifts on the fitted value of \GW in the combined PLH fits, using the (a) \ptl and (b) \mt distributions and the CT18 PDF set. For a given NP $\theta$, the shift is defined as the product of its post-fit value $\hat \theta$ and its pre-fit impact on \GW. The points, which are plotted according to the bottom horizontal scale, show $\hat \theta$ for each of the nuisance parameters. The error bars show the corresponding post-fit uncertainties, $\sigma_{\hat\theta}$. The nuisance parameters are ranked according to the shift induced on \GW, the NPs with the largest shifts at the top.}\label{fig:gwshiftrank}
\end{figure}

Figure~\ref{fig:gwshiftrank} summarises the ten nuisance parameters that induce the largest shift of \GW in fits to the \ptl and \mt distributions. The largest shifts are related to the multijet~(MJ) background, to the lepton calibration, to specific eigenvectors of the CT18 PDF set, to the luminosity, and to the uncertainty in charm-induced production for the \ptw description.

The decomposition of post-fit uncertainties is done with the same method as in the \mW measurement, see Section~\ref{sec:mWcomb}. A summary of the uncertainties contributed by various sources is given in Table~\ref{tab:gwunc}. The measurement is dominated by systematic uncertainties for the \ptl\ distribution, while for the \mt\ distribution statistical and systematic uncertainties are of similar magnitude. The dominant systematic uncertainties are due to the parton shower modelling for \ptl, and lepton and recoil performance for \mt, respectively.

\begin{table}[tbp]
\center
\caption{Uncertainty components for the \ptl, \mt and combined \GW measurements using the CT18 PDF set. The first columns give the total, statistical and overall systematic uncertainty in the measurements. The following columns show the contributions of modelling and experimental systematic uncertainties, grouped into categories.\label{tab:gwunc}}
\resizebox{\textwidth}{!}{\begin{tabular}{l|ccc|cccccccccc}
\toprule
Unc. [\MeV] & Total  & Stat. & Syst. & PDF & $A_i$ & Backg. & EW & $e$ & $\mu$ & \ut & Lumi & \mW & PS \\
\midrule
\ptl   & 72  & 27 & 66  & 21  & 14  & 10  & 5   & 13  & 12  & 12  & 10   & 6   & 55 \\
\mt    & 48  & 36 & 32  & 5   & 7   & 10   & 3   & 13  & 9   & 18  & 9   & 6   & 12 \\
Combined & 47  & 32  & 34  & 7   & 8   & 9   & 3   & 13  & 9   & 17  & 9   & 6   & 18 \\
\bottomrule
\end{tabular}}
\end{table}%

\begin{figure}[tbp]
\centering
\subfigure[]{\includegraphics[width=0.48\textwidth]{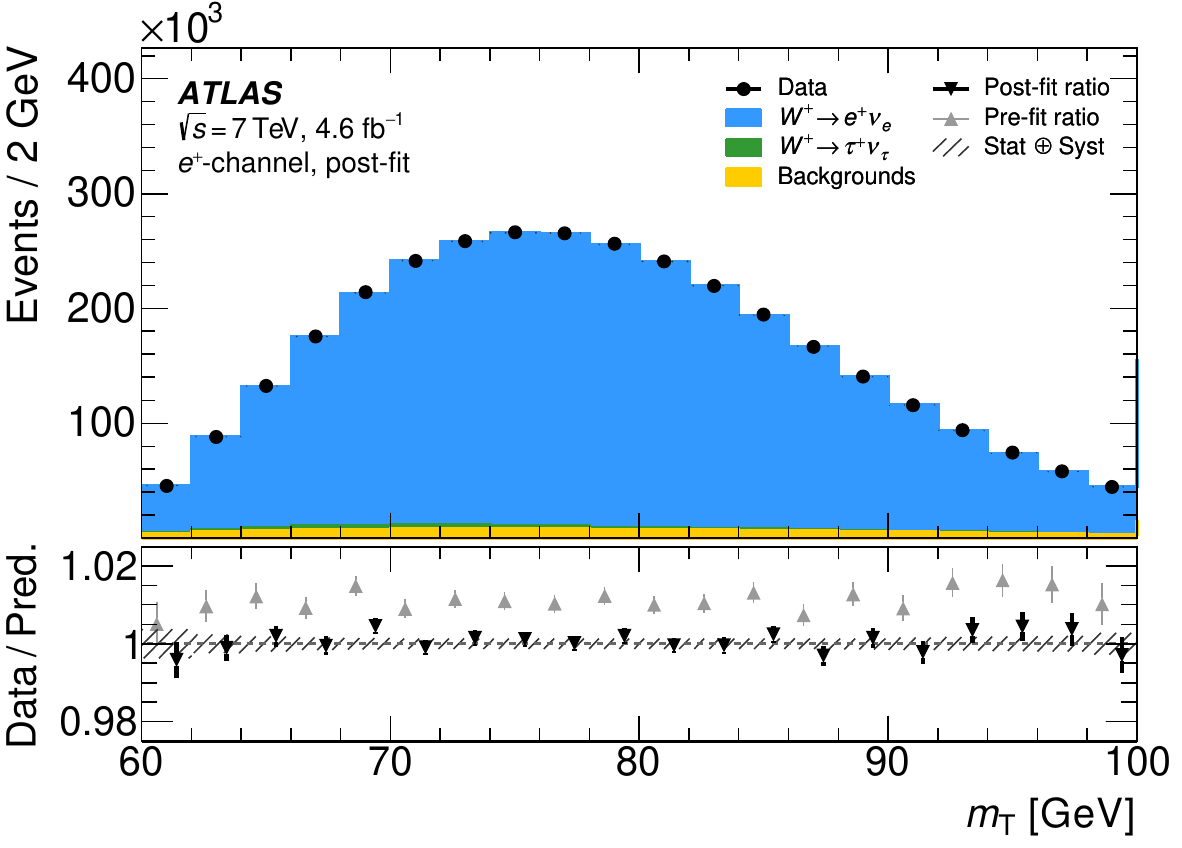}}
\subfigure[]{\includegraphics[width=0.48\textwidth]{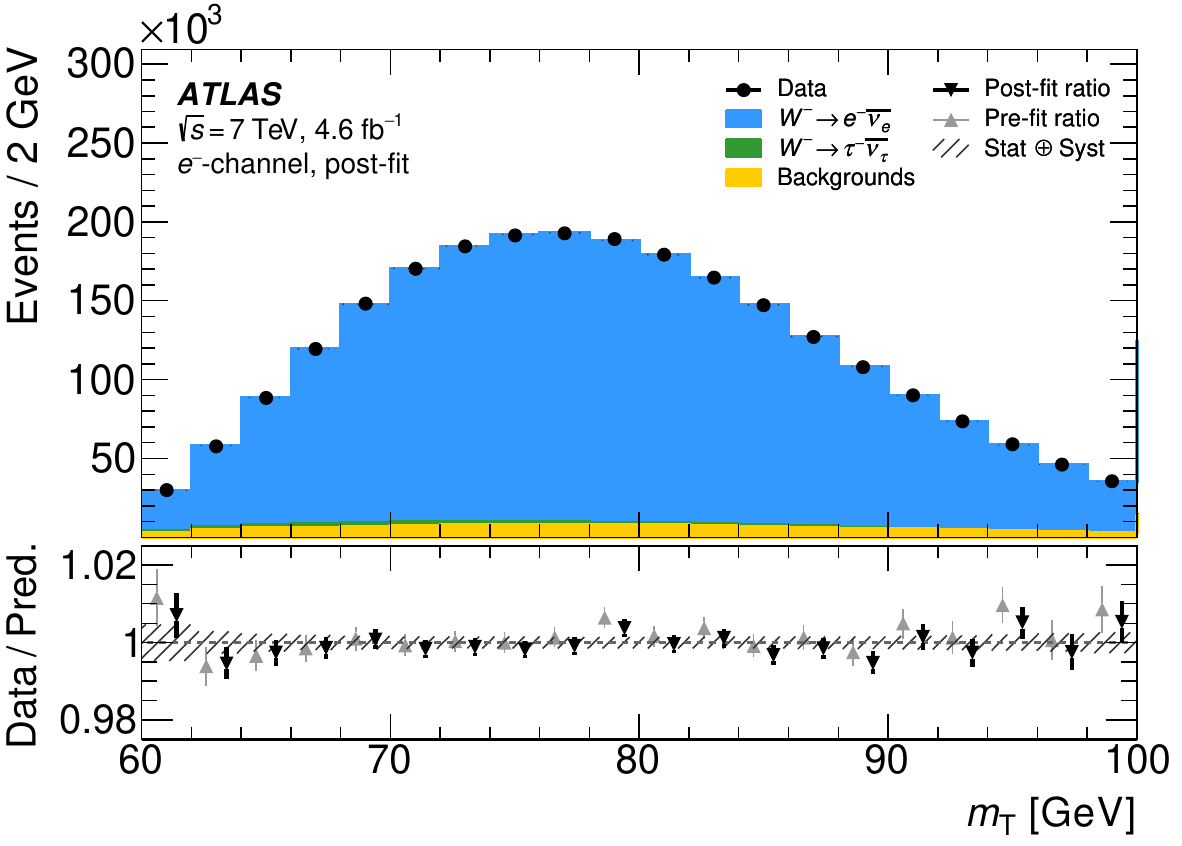}}
\subfigure[]{\includegraphics[width=0.48\textwidth]{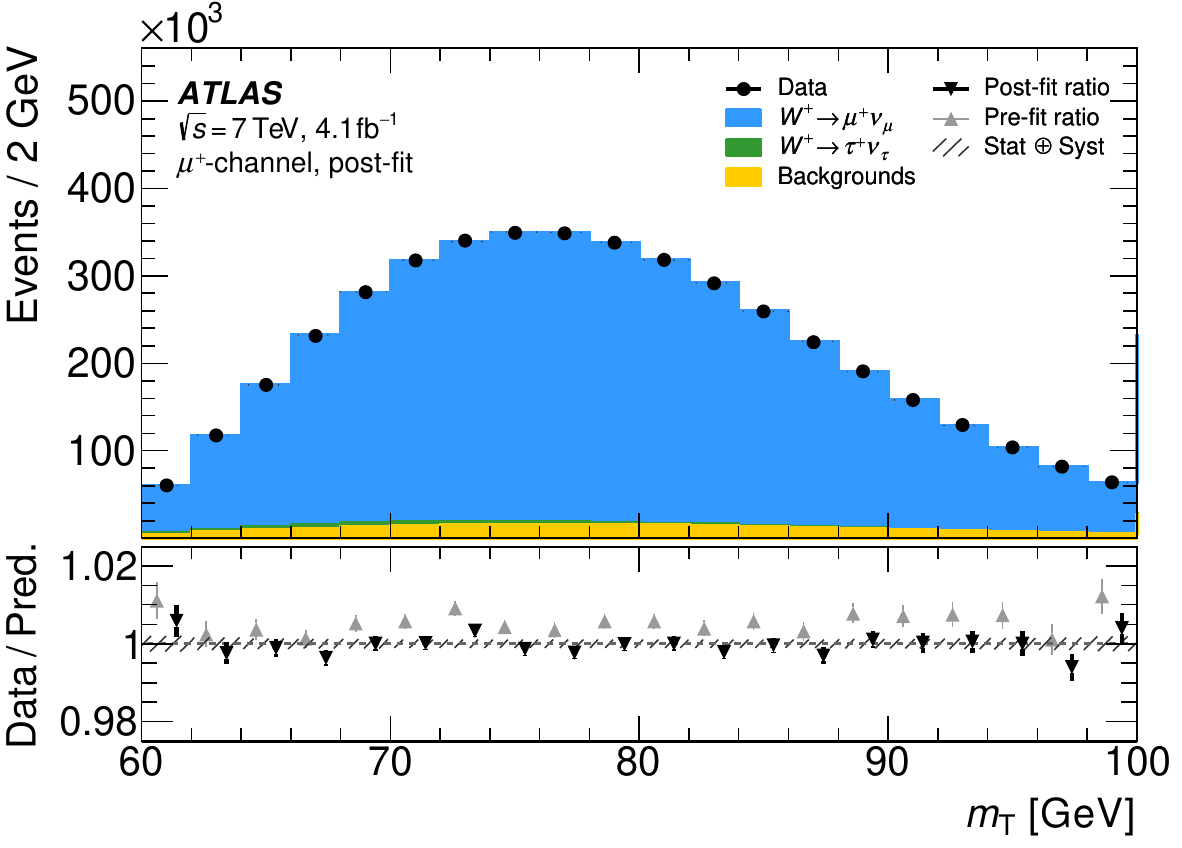}}
\subfigure[]{\includegraphics[width=0.48\textwidth]{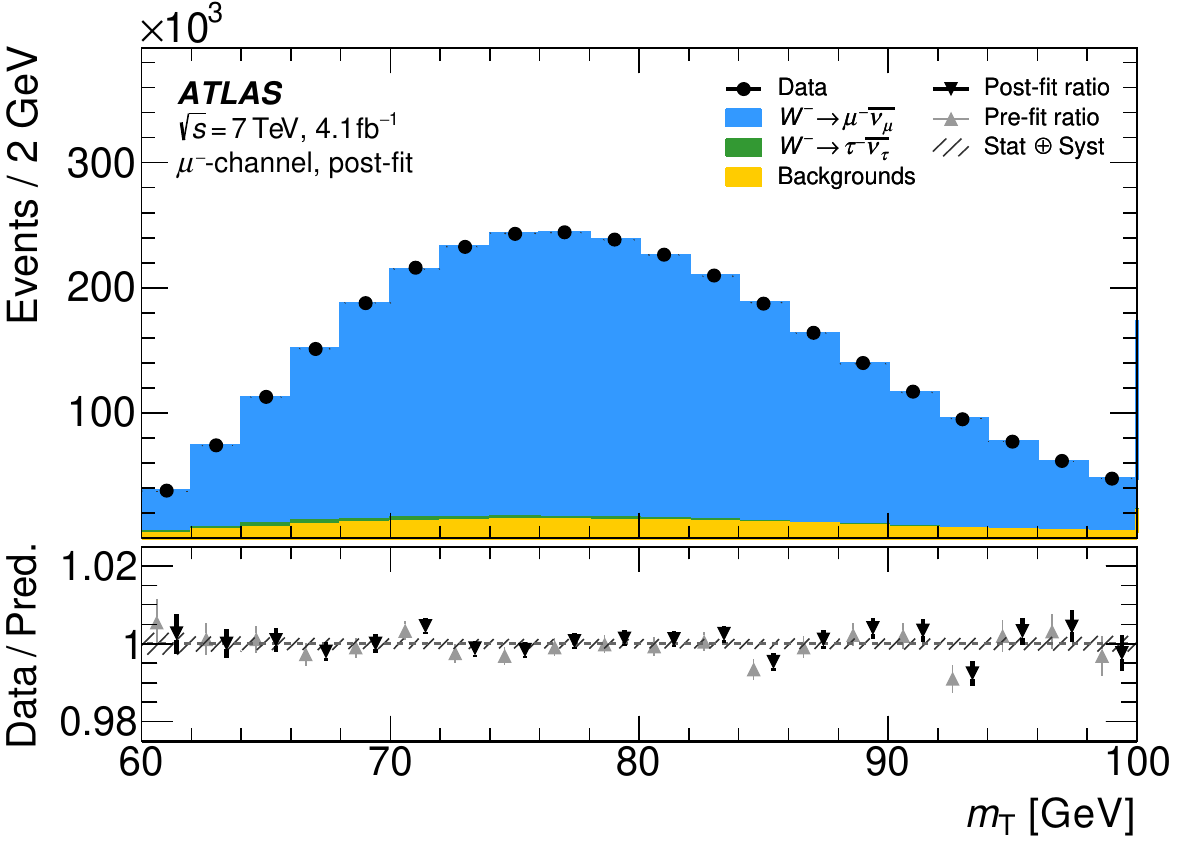}}
\caption{Post-fit distributions of \mt with data and MC for (a) $W^+\rightarrow e^+\nu_e$, (b) $W^-\rightarrow e^-\overline{\nu}_e$, (c) $W^+\rightarrow \mu^+\nu_{\mu}$ and (d) $W^-\rightarrow \mu^-\overline{\nu}_{\mu}$, inclusive over all $\eta$ regions, and using the CT18 PDF set. In the bottom panels, the darker points represent the post-fit ratio of data to MC, while the lighter points indicate the ratio before the fit. The hatched band represents the total uncertainty of the data.}
\label{fig:gwpostfitPlots}
\end{figure}

An overview of selected pre- and post-fit distributions of \mt is shown in Figure~\ref{fig:gwpostfitPlots}, where a general better agreement can be observed for the post-fit case. The post-fit distributions use the final measured value of \GW.

\subsection{Combination}

The combination of results obtained from \ptl and \mt distributions follows the procedure described in Section~\ref{sec:mWcomb}. The \ptl and \mt results are fully compatible with each other in terms of central values and nuisance parameters. The results for all considered PDF sets are presented in Table~\ref{tab:combigw}. The weight of the \mt fit ranges from 84\% to 89\%, depending on the PDF set, and dominates the final result. For CT18, the final result yields:

$$\GW = 2202 \pm 32~(\text{stat.}) \pm 34~(\text{syst.}) ~\MeV = 2202 \pm 47 ~\MeV,$$

where the first uncertainty component is statistical and the second corresponds to the total systematic uncertainties.
The compatibility of the measured value with the SM expectation is illustrated in Figure~\ref{fig:WWidthFinalResults}, together with selected previous measurements.

\begin{table}[tbp]
\center
\caption{Uncertainty correlation between the \ptl and \mt fits, combination weights and combination results for $\GW$ and the indicated PDF sets.\label{tab:combigw}}
\begin{tabular}{l|cccc}
\toprule
PDF set & Correlation & weight (\mt) & weight (\ptl) & Combined $\GW$ [\MeV] \\
\midrule
CT14       & 50.3\% &  88\% & 12\% & $2204 \pm 47$ \\
CT18       & 51.5\% &  87\% & 13\% & $2202 \pm 47$ \\
CT18A      & 50.0\% &  86\% & 14\% & $2184 \pm 47$ \\
MMHT2014   & 50.8\% &  88\% & 13\% & $2182 \pm 47$ \\
MSHT20     & 53.6\% &  89\% & 11\% & $2181 \pm 47$ \\
ATLASpdf21 & 49.5\% &  84\% & 16\% & $2193 \pm 46$ \\
NNPDF31    & 49.9\% &  86\% & 14\% & $2182 \pm 46$ \\
NNPDF40    & 51.4\% &  85\% & 15\% & $2184 \pm 46$ \\
\bottomrule
\end{tabular}
\end{table}

\FloatBarrier



%
\section{Simultaneous determination of the $W$-boson mass and width }
The previously described determination of \mW assumes for the $W$-boson width its SM value and uncertainty, and similarly the \GW measurement uses the SM prediction for \mW. To further test the interplay between the two observables, the PLH fit is also performed with both \mW and \GW free in the fit. This fit with two parameters of interest relies on the same experimental calibrations and physics modelling.
The fit yields values of $\mW = 80351.8 \pm 16.7$~\MeV and $\GW = 2216 \pm 73$~\MeV for the \ptl distributions and $\mW = 80369.4 \pm 26.8$~\MeV and $\GW = 2186 \pm 53$~\MeV for the \mt distributions using the CT18 PDF set. Compared with the separate \mW and \GW fits, the decrease of \mW by 10~\MeV and 25~\MeV for the \ptl and \mt distributions, respectively, is consistent with the larger measured value of \GW following the observed anti-correlation of \mW and \GW. The increase in total uncertainty is due to the removal of the constraints on \GW and \mW in the fit.

The combination of results obtained from \ptl and \mt distributions follows the procedure described in Section~\ref{sec:mWcomb}. The \ptl and \mt results are fully compatible with each other.

\begin{figure}[tbp]
\centering
\subfigure[]{\includegraphics[width=0.49\linewidth]{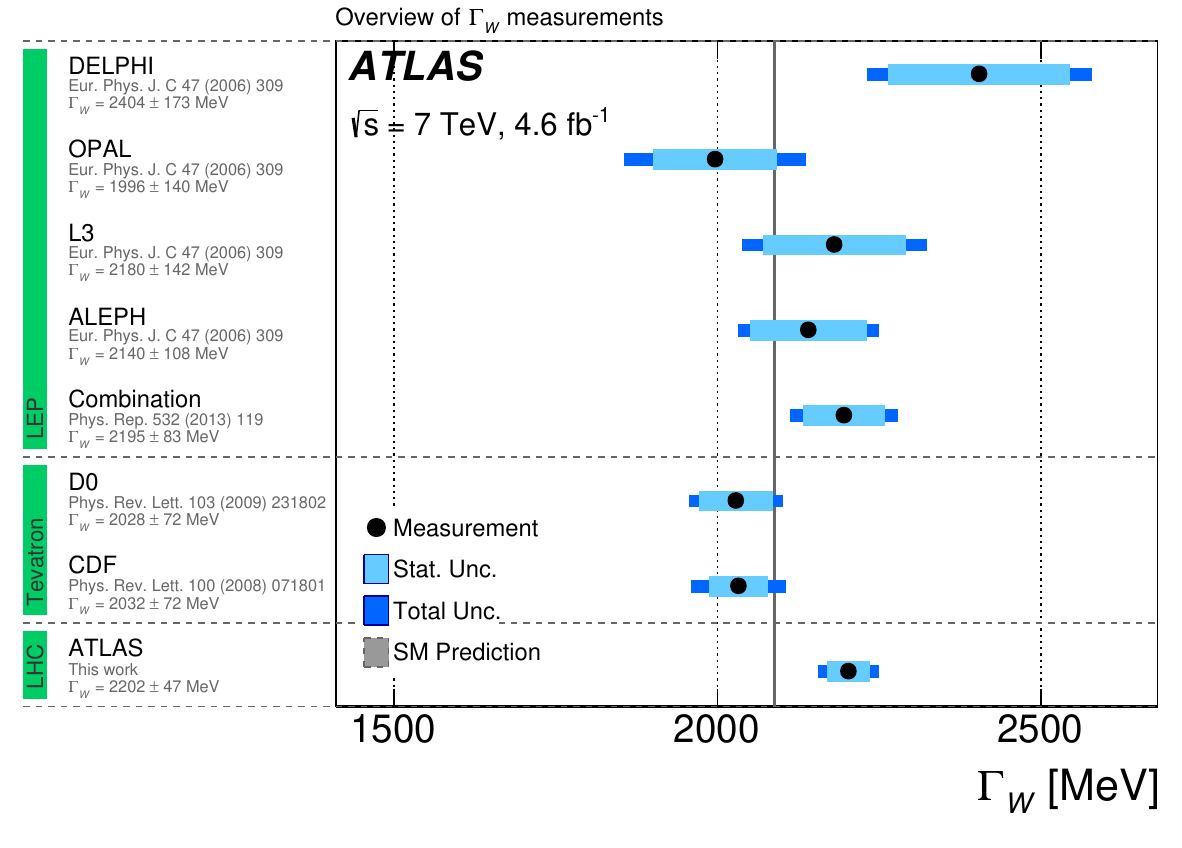}\label{fig:WWidthFinalResults}}
\subfigure[]{\includegraphics[width=0.49\textwidth]{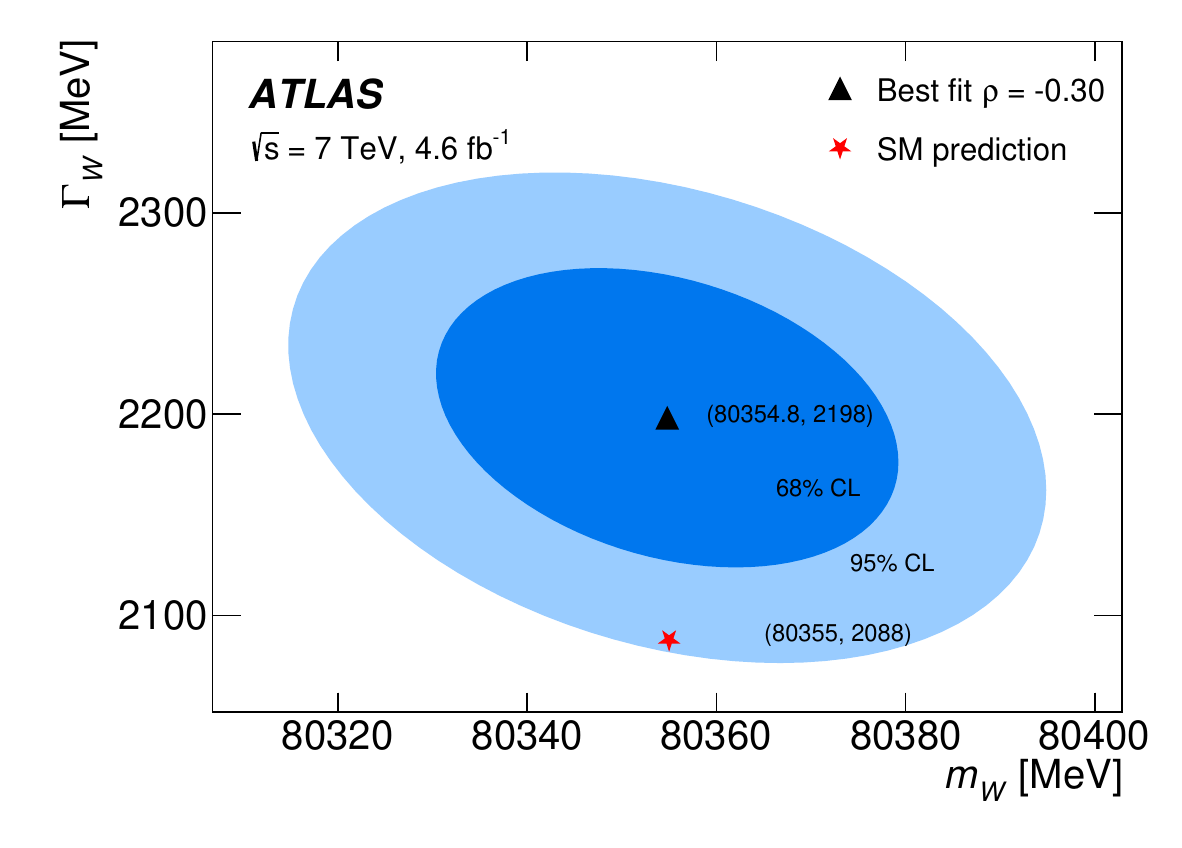}\label{fig:mwgwellipses}}
\caption{(a) Present measurement of \GW, compared to the SM prediction from the global electroweak fit~\cite{deBlas:2021wap}, and to the measurements of LEP~\cite{ALEPH:2013dgf} and Tevatron~\cite{thetevatronelectroweakworkinggroup2010combination}. (b) 68\% and 95\% CL uncertainty contours for the simultaneous determination of \mW and \GW using the CT18 PDF set and combining results from the \ptl and \mt distributions. The triangular marker represents the best fit, while the star corresponds to the SM prediction of Ref.~\cite{deBlas:2021wap}.}
\end{figure}

For the CT18 PDF set, the combination yields values of

$$\mW = 80354.8 \pm 16.1 ~\MeV$$

and

$$\GW = 2198 \pm 49 ~\MeV,$$

with a correlation of $-30$\% that reflects the negative slope of the dependencies reported in Sections~\ref{sec:mWcomb} and~\ref{sec:gWresults}. The 68\% and 95\% CL uncertainty contours are shown in Figure~\ref{fig:mwgwellipses}.

%
\section{Conclusion}

This paper reports on a first measurement of the $W$-boson width at the LHC as well as the reanalysis of the data used in the published $W$-boson mass measurement, using an improved fitting technique and updated parton distribution functions. Both measurements are based on proton--proton collision data at a centre-of-mass energy of $\sqrt{s} = 7~\TeV$ recorded by the ATLAS detector at the LHC in 2011, and corresponding to an integrated luminosity of 4.6~fb$^{-1}$ and $4.1$~fb$^{-1}$ in the electron and muon channels, respectively.

The measurements of \mW using the \ptl and \mt distributions are found to be consistent and their combination yields
\begin{eqnarray}
\nonumber \mW = 80366.5 \pm 9.8~(\text{stat.}) \pm 12.5~(\text{syst.})~\MeV = 80366.5 \pm 15.9~\MeV.
\end{eqnarray}
The present result is compatible with and supersedes the previous measurement of \mW at ATLAS using the same data. No significant deviation from the SM expectation is observed.
The PDF dependence of the \mW result is driven by the pre-fit PDF uncertainties, and is strongly reduced when allowing for enlarged uncertainties. The final results are obtained using the CT18 PDF set, which is the most conservative PDF set for these measurements and compatible with the fits using enlarged PDF uncertainties of other sets.

The measurements of \GW using the \ptl and \mt distributions are also found to be consistent and their combination yields a value of
\begin{eqnarray}
\nonumber \GW = 2202 \pm 32~(\text{stat.}) \pm 34~(\text{syst.})~\MeV = 2202 \pm 47 ~\MeV.
\end{eqnarray}
It is the currently most precise single measurement of \GW and agrees with the SM expectation of $\GW^\text{SM} = 2088 \pm 1$~\MeV within 2.4 standard deviations. The dependence of \GW on the assumed PDF set is weak compared with the measurement uncertainties.


%
\section*{Acknowledgements}
%

%
%

%
%

We thank CERN for the very successful operation of the LHC and its injectors, as well as the support staff at
CERN and at our institutions worldwide without whom ATLAS could not be operated efficiently.

The crucial computing support from all WLCG partners is acknowledged gratefully, in particular from CERN, the ATLAS Tier-1 facilities at TRIUMF/SFU (Canada), NDGF (Denmark, Norway, Sweden), CC-IN2P3 (France), KIT/GridKA (Germany), INFN-CNAF (Italy), NL-T1 (Netherlands), PIC (Spain), RAL (UK) and BNL (USA), the Tier-2 facilities worldwide and large non-WLCG resource providers. Major contributors of computing resources are listed in Ref.~\cite{ATL-SOFT-PUB-2023-001}.

We gratefully acknowledge the support of ANPCyT, Argentina; YerPhI, Armenia; ARC, Australia; BMWFW and FWF, Austria; ANAS, Azerbaijan; CNPq and FAPESP, Brazil; NSERC, NRC and CFI, Canada; CERN; ANID, Chile; CAS, MOST and NSFC, China; Minciencias, Colombia; MEYS CR, Czech Republic; DNRF and DNSRC, Denmark; IN2P3-CNRS and CEA-DRF/IRFU, France; SRNSFG, Georgia; BMBF, HGF and MPG, Germany; GSRI, Greece; RGC and Hong Kong SAR, China; ISF and Benoziyo Center, Israel; INFN, Italy; MEXT and JSPS, Japan; CNRST, Morocco; NWO, Netherlands; RCN, Norway; MNiSW, Poland; FCT, Portugal; MNE/IFA, Romania; MESTD, Serbia; MSSR, Slovakia; ARRS and MIZ\v{S}, Slovenia; DSI/NRF, South Africa; MICINN, Spain; SRC and Wallenberg Foundation, Sweden; SERI, SNSF and Cantons of Bern and Geneva, Switzerland; MOST, Taipei; TENMAK, T\"urkiye; STFC, United Kingdom; DOE and NSF, United States of America.

Individual groups and members have received support from BCKDF, CANARIE, CRC and DRAC, Canada; PRIMUS 21/SCI/017, CERN-CZ and FORTE, Czech Republic; COST, ERC, ERDF, Horizon 2020, ICSC-NextGenerationEU and Marie Sk{\l}odowska-Curie Actions, European Union; Investissements d'Avenir Labex, Investissements d'Avenir Idex and ANR, France; DFG and AvH Foundation, Germany; Herakleitos, Thales and Aristeia programmes co-financed by EU-ESF and the Greek NSRF, Greece; BSF-NSF and MINERVA, Israel; Norwegian Financial Mechanism 2014-2021, Norway; NCN and NAWA, Poland; La Caixa Banking Foundation, CERCA Programme Generalitat de Catalunya and PROMETEO and GenT Programmes Generalitat Valenciana, Spain; G\"{o}ran Gustafssons Stiftelse, Sweden; The Royal Society and Leverhulme Trust, United Kingdom.

In addition, individual members wish to acknowledge support from CERN: European Organization for Nuclear Research (CERN PJAS); Chile: Agencia Nacional de Investigaci\'on y Desarrollo (FONDECYT 1190886, FONDECYT 1210400, FONDECYT 1230812, FONDECYT 1230987); China: Chinese Ministry of Science and Technology (MOST-2023YFA1605700), National Natural Science Foundation of China (NSFC - 12175119, NSFC 12275265, NSFC-12075060); Czech Republic: Czech Science Foundation (GACR - 24-11373S), Ministry of Education Youth and Sports (FORTE CZ.02.01.01/00/22\_008/0004632), PRIMUS Research Programme (PRIMUS/21/SCI/017); EU: H2020 European Research Council (ERC - 101002463); European Union: European Research Council (ERC - 948254, ERC 101089007), Horizon 2020 Framework Programme (MUCCA - CHIST-ERA-19-XAI-00), European Union, Future Artificial Intelligence Research (FAIR-NextGenerationEU PE00000013), Italian Center for High Performance Computing, Big Data and Quantum Computing (ICSC, NextGenerationEU); France: Agence Nationale de la Recherche (ANR-20-CE31-0013, ANR-21-CE31-0013, ANR-21-CE31-0022), Investissements d'Avenir Labex (ANR-11-LABX-0012); Germany: Baden-Württemberg Stiftung (BW Stiftung-Postdoc Eliteprogramme), Deutsche Forschungsgemeinschaft (DFG - 469666862, DFG - CR 312/5-2); Italy: Istituto Nazionale di Fisica Nucleare (ICSC, NextGenerationEU); Japan: Japan Society for the Promotion of Science (JSPS KAKENHI JP21H05085, JSPS KAKENHI JP22H01227, JSPS KAKENHI JP22H04944, JSPS KAKENHI JP22KK0227); Netherlands: Netherlands Organisation for Scientific Research (NWO Veni 2020 - VI.Veni.202.179); Norway: Research Council of Norway (RCN-314472); Poland: Polish National Agency for Academic Exchange (PPN/PPO/2020/1/00002/U/00001), Polish National Science Centre (NCN 2021/42/E/ST2/00350, NCN OPUS nr 2022/47/B/ST2/03059, NCN UMO-2019/34/E/ST2/00393, UMO-2020/37/B/ST2/01043, UMO-2021/40/C/ST2/00187, UMO-2022/47/O/ST2/00148, UMO-2023/49/B/ST2/04085); Slovenia: Slovenian Research Agency (ARIS grant J1-3010); Spain: Generalitat Valenciana (Artemisa, FEDER, IDIFEDER/2018/048), Ministry of Science and Innovation (MCIN \& NextGenEU PCI2022-135018-2, MICIN \& FEDER PID2021-125273NB, RYC2019-028510-I, RYC2020-030254-I, RYC2021-031273-I, RYC2022-038164-I), PROMETEO and GenT Programmes Generalitat Valenciana (CIDEGENT/2019/023, CIDEGENT/2019/027); Sweden: Swedish Research Council (Swedish Research Council 2023-04654, VR 2018-00482, VR 2022-03845, VR 2022-04683, VR 2023-03403, VR grant 2021-03651), Knut and Alice Wallenberg Foundation (KAW 2018.0157, KAW 2018.0458, KAW 2019.0447, KAW 2022.0358); Switzerland: Swiss National Science Foundation (SNSF - PCEFP2\_194658); United Kingdom: Leverhulme Trust (Leverhulme Trust RPG-2020-004), Royal Society (NIF-R1-231091); United States of America: U.S. Department of Energy (ECA DE-AC02-76SF00515), Neubauer Family Foundation.

%
%


%
%
%
%
%
%
%

%
%
%
%
%
%

%
%

%
%
%
\printbibliography
\clearpage
 
\begin{flushleft}
\hypersetup{urlcolor=black}
{\Large The ATLAS Collaboration}

\bigskip

\AtlasOrcid[0000-0002-6665-4934]{G.~Aad}$^\textrm{\scriptsize 103}$,
\AtlasOrcid[0000-0001-7616-1554]{E.~Aakvaag}$^\textrm{\scriptsize 16}$,
\AtlasOrcid[0000-0002-5888-2734]{B.~Abbott}$^\textrm{\scriptsize 121}$,
\AtlasOrcid[0000-0002-0287-5869]{S.~Abdelhameed}$^\textrm{\scriptsize 117a}$,
\AtlasOrcid[0000-0002-1002-1652]{K.~Abeling}$^\textrm{\scriptsize 55}$,
\AtlasOrcid[0000-0001-5763-2760]{N.J.~Abicht}$^\textrm{\scriptsize 49}$,
\AtlasOrcid[0000-0002-8496-9294]{S.H.~Abidi}$^\textrm{\scriptsize 29}$,
\AtlasOrcid[0009-0003-6578-220X]{M.~Aboelela}$^\textrm{\scriptsize 44}$,
\AtlasOrcid[0000-0002-9987-2292]{A.~Aboulhorma}$^\textrm{\scriptsize 35e}$,
\AtlasOrcid[0000-0001-5329-6640]{H.~Abramowicz}$^\textrm{\scriptsize 152}$
% ...
\bigskip
\\

% ...
$^{16}$Department for Physics and Technology, University of Bergen, Bergen; Norway.\\
% ...
$^{29}$Physics Department, Brookhaven National Laboratory, Upton NY; United States of America.\\
% ...
$^{35}$$^{(a)}$Facult\'e des Sciences Ain Chock, R\'eseau Universitaire de Physique des Hautes Energies - Universit\'e Hassan II, Casablanca;$^{(b)}$Facult\'{e} des Sciences, Universit\'{e} Ibn-Tofail, K\'{e}nitra;$^{(c)}$Facult\'e des Sciences Semlalia, Universit\'e Cadi Ayyad, LPHEA-Marrakech;$^{(d)}$LPMR, Facult\'e des Sciences, Universit\'e Mohamed Premier, Oujda;$^{(e)}$Facult\'e des sciences, Universit\'e Mohammed V, Rabat;$^{(f)}$Institute of Applied Physics, Mohammed VI Polytechnic University, Ben Guerir; Morocco.\\
% ...
$^{44}$Physics Department, Southern Methodist University, Dallas TX; United States of America.\\
% ...
$^{49}$Fakult\"{a}t Physik , Technische Universit{\"a}t Dortmund, Dortmund; Germany.\\
% ...
$^{55}$II. Physikalisches Institut, Georg-August-Universit\"{a}t G\"ottingen, G\"ottingen; Germany.\\
% ...
$^{103}$CPPM, Aix-Marseille Universit\'e, CNRS/IN2P3, Marseille; France.\\
% ...
$^{117}$$^{(a)}$New York University Abu Dhabi, Abu Dhabi;$^{(b)}$United Arab Emirates University, Al Ain; United Arab Emirates.\\
% ...
$^{121}$Homer L. Dodge Department of Physics and Astronomy, University of Oklahoma, Norman OK; United States of America.\\
% ...
$^{152}$Raymond and Beverly Sackler School of Physics and Astronomy, Tel Aviv University, Tel Aviv; Israel.\\
% ...
$^{a}$ Also Affiliated with an institute covered by a cooperation agreement with CERN.\\
$^{b}$ Also at An-Najah National University, Nablus; Palestine.\\
$^{c}$ Also at Borough of Manhattan Community College, City University of New York, New York NY; United States of America.\\
$^{d}$ Also at Center for Interdisciplinary Research and Innovation (CIRI-AUTH), Thessaloniki; Greece.\\
$^{e}$ Also at Centro Studi e Ricerche Enrico Fermi; Italy.\\
$^{f}$ Also at CERN, Geneva; Switzerland.\\
% ...

\end{flushleft}


%

%
%

%

%
%
%
%

\end{document}